\documentclass[10pt, preprint]{aastex}

\usepackage{graphics}

\usepackage{amsmath}

\usepackage{color}
\def\tcb{\textcolor{blue}}
\def\tcr{\textcolor{red}}
\usepackage{epstopdf}

\newcommand{\bea}{\begin{eqnarray}}
\newcommand{\eea}{\end{eqnarray}}
\def\beq{\begin{equation}}
\def\eeq{\end{equation}}

\def\Ms{$M_\odot$}
\def\Ls{$L_\odot$}
\def\Rs{$R_\odot$\xspace}
\def\Angst{$\buildrel _{\circ} \over {\mathrm{A}}$}
\def\de{$^{\circ}$\xspace}
\def\rel{relativistic \,}
\def\pwt{{PWT}\,}
\def\r{{\it RHESSI}\ }
\def\f{{\it Fermi}\ }
\def\min{{\rm min}}
\def\max{{\rm max}}
\def\ie{{\it i.e.\ }}
\def\ea{{\it et al.\ }}
\def\eg{{\it e.g.\ }}

\def\a{{\alpha}}
\def\b{{\beta}}
\def\d{{\delta}}
\def\D{{\Delta}}
\def\g{{\gamma}}
\def\G{\Gamma}
\def\e{\epsilon}
\def\l{{\lambda}}
\def\L{{\Lambda}}
\def\t{{\theta}}

\def\cl{\centerline}
\def\ni{\noindent}
\def\ss{\smallskip}
\def\bs{\bigskip}
\def\ms{\medskip}
\def\np{\newpage}
\def\rs{{R_\odot}} 

\def\tsc{\tau_{\rm sc}\,}
\def\tac{\tau_{\rm ac}\,}
\def\tdiff{\tau_{\rm diff}\,}
\def\tcross{\tau_{\rm cross}\,}
\def\tesc{T_{\rm esc}\,}
\def\tescu{T^{\rm up}_{\rm esc}\,}
\def\tescd{T^{\rm down}_{\rm esc}\,}
\def\tloss{\tau_{\rm L}\,}

\def\ni{\noindent}
\def\ss{\smallskip}
\def\bs{\bigskip}
\def\ms{\medskip}
\def\np{\newpage}

\shorttitle{Cosmic Ray Electron  Transport in Inner Helisphere}
\shortauthors{Petrosian, Orlando, \& Strong}

\begin{document}
	
	\title{TRANSPORT OF COSMIC RAY ELECTRONS FROM 1 AU TO THE SUN}
	
	\author{Vah\'e Petrosian}
	
	\affil{Department of Physics  Stanford University, 382 Via Pueblo Mall, Stanford, CA 94305-4060}
	\affil{Kavli Institute for Particle Astrophysics and Cosmology, Stanford University}
	\affil{Also Department of Applied Physics, Stanford University}
	
	\author{Elena Orlando}

	\affil{Department of Physics, University of Trieste,  Trieste, Italy}
	\affil{Also Kavli Institute for Particle Astrophysics and Cosmology, Stanford University}
	\affil{National Institute for Nuclear Physics (INFN) Trieste, Italy}
	
	\author{Andrew Strong}
	
	\affil{Max-Planck-Institut für extraterrestrische Physik, Garching, Germany}
	
	\email{vahep@stanford.edu}
	
	\begin{abstract}
		
		Gamma rays  are produced by cosmic ray (CR)  protons  interacting with the  particles at  solar
		photosphere and by cosmic ray  electrons and positrons (CRes) via inverse Compton   scattering of
		solar photons. The former come from the solar disk while the latter extend beyond the disk.
		Evaluation of these emissions requires the flux and spectrum of CRs in the vicinity of the Sun,
		while most observations provide flux and spectra near the Earth, at around 1 AU from the Sun. Past
		estimates of the quiet Sun gamma-ray emission use phenomenological modulation procedures to estimate
		spectra near the Sun (see review by \citet{Orlando} and references therein). We show that CRe transport in the inner heliosphere requires a kinetic approach  and use a novel approximation to determine the
		variation of CRe flux and spectrum from 1 AU to the Sun including  effects of (1) the structure of 
		large scale magnetic field,  (2) small scale turbulence in the solar wind from several  in situ measurements, in particular, those by Parker Solar Probe that
		extend this information to 0.1 AU, and (3) most importantly,   energy losses due
		to synchrotron  and inverse  Compton processes.  We present results on the flux and spectrum variation of CRes from 1 AU to the Sun for several transport models. In  forthcoming
		papers we will use these results for a more accurate estimate of quiet Sun inverse Compton gamma-ray spectra,
		and, for the first time, the spectrum of extreme ultraviolet  to hard X-ray  photons produced by
		synchrotron emission. These  can be compared with the quiet Sun gamma-ray observation by \f (see, e.g.~\f-LAT Collaboration, 2011) and X-ray upper limits set by \r (Hannah et al., 2010). 
		
	\end{abstract}
	
	\keywords{Cosmic Rays; transport of particles--Solar Wind: --Sun: particle emissions
		--turbulence}
	
	\section{Introduction}
	\label{sec:intro}

	Spectra and many other characteristics of high energy cosmic rays (CRs) have been directly observed
	and investigated for more than a century by various instruments. These characteristics can be also
	deduced by the radiation they produce interacting with the diffuse interstellar particles, photons and
	magnetic fields; gamma-rays from decay of pions  produced by interaction of CR ions (mostly protons;
	CRps) and from inverse Compton (IC) scattering of low energy photons (mainly starlight) of CR
	electrons and positrons (CRes), and radio radiation produced by CRes via  synchrotron mechanisms.
	Similar radiation can be produce by the interactions of CRs with denser objects like stars, planets
	and satellites. 
	{\it EGRET} on board Compton Gamma-Ray Observatory (CGRO) was first to detect gamma-rays from the  quiet phase of the Sun (QS) \citep{Orlando2008}. Over the past decade the  Large Area Telescope (LAT) on board \f has provided a rich body of data on
	$>100$ MeV gamma-rays during QS, which consist of disk
	emission due to pion decay and somewhat extended emission due to the IC scattering of solar
	optical photons by CRes.  These observations  have been investigated  extensively, commonly using  a phenomenological description of  solar modulation of the CRs (see, e.g.~\citet{Abdo11}, \citet{Fujii}, \citet{Moskalenko} and \citet{Orlando2007}). 
	
	However, to the best of our knowledge, there
	has not  been much discussion, or any detailed analysis, of the synchrotron emission by CRes.
	Evaluation of the synchrotron emission
	during active phase of the sun with many active regions and strong complex magnetic field structure
	is complicated. But during QS periods the magnetic field  in the heliosphere
	from  photosphere to 1 AU varies fairly smoothly (approximately following the Parker spiral
	structure) from $B\sim 10$ G to tens of $\mu$G (or few nT). Thus,
	GeV to TeV CRes can produce synchrotron radiation from few GHz to $10^{15}$ Hz at 1 AU ($B\sim 4$
	nT) and from $\sim 10^{15}$
	Hz ($\sim 1$ eV) to $\sim 10^{21}$ Hz ($\sim $ MeV) near the photosphere ($B\sim 10$ G). Most of this
	radiation will be
	undetectable or fall below the radiation produced by other mechanisms. However, recent analysis of
	the \r observation of the Sun (Hannah et al. 2010) during the QS phase show
	some robust   upper limits on the flux in the hard X-ray (HXR) range. 
	
	Our ultimate goal  is to
	investigate the possibility of
	detecting synchrotron radiation during the transport of CRes from 1 AU to the Sun and test whether
	the observed QS HXR upper limits can constrain this model. This requires an accurate determination
	of the spectral variation of the CRes from 1 AU, where they are observed, to the Sun. As mentioned above, past works have used phenomenological modulation approach, application of which to the inner heliosphere is highly uncertain. As we will show, this tasks requires a kinetic approach, which we develop in this paper. The
	result from such a study can also provide a more accurate determination of the expected IC gamma-ray
	emission. 
	The focus of the current paper is  the transport of CRes from 1 AU to the Sun. In  subsequent papers
	we will address the emission characteristics.

	In the next section we describe several ingredients that are needed for calculation of the spectrum
	of the synchrotron and IC emission from CRes during their transport through the {\it inner heliosphere} from 1 AU to the
	Sun. In \S 3 we discuss the coefficients of the transport kinetic equation  and in \S 4 we calculate
	the CRe spectral variation for three models, and present equation for evaluation of radiation spectra that can be observed at 1 AU.  A brief summary and conclusions are presented in \S 5.

	\section{Synchrotron and IC Emissivity}
	\label{sec:emissivities}

	The mono-energetic  spectral emissions  of  relativistic
	electrons (mass $m_e$, charge $e$) with
	Lorentz factor $\g$ and pitch angle $\a$ (or its cosine $\mu=\cos \a$) at a distance $r$ from the center of the Sun can be described by the general function $k(\nu, \g, \mu, r)$ (in ergs  s$^{-1}$ Hz$^{-1}$), which varies with $r$ because of the variation of the magnetic field, $B(r)$  (for synchrotron), and photon energy density, $u_{\rm ph}(r)$ (for IC).
	
	The
	emissivity (in ergs s$^{-1}$ Hz$^{-1}$ cm$^{-3}$) of a population of electrons is obtained by integrating over the electron energy (or Lorentz
	factor)  and pitch angle  distribution,  
	$N(\g, \mu, r)$ (in cm$^{-3}$  $\g^{-1}$, rad$^{-1}$), as
	\beq
	\label{emissivity}
	\eta(\nu, r)=\int_{-1}^1 d\mu\int_{\g_{\rm min}}^\infty N(\g, \mu, r)k(\nu,\g,\mu,r)d\g, 
	\eeq
	where  $\g_{\rm min}\gg 1$ is the lowest observed Lorentz factor.
	
	The {\it two main ingredients} needed for evaluation of the emissivity are the variations of magnetic field,
	$B(r)$, and optical photon energy density, $u_{\rm ph}(r),$
	and  energy and pitch angle distribution of the CRes with distance
	from the Sun.

	\subsection{Structure of the Magnetic Field}
	\label{sec:bfield}
	
\begin{figure}[!ht]
		\begin{center}
			\includegraphics[width=3in]{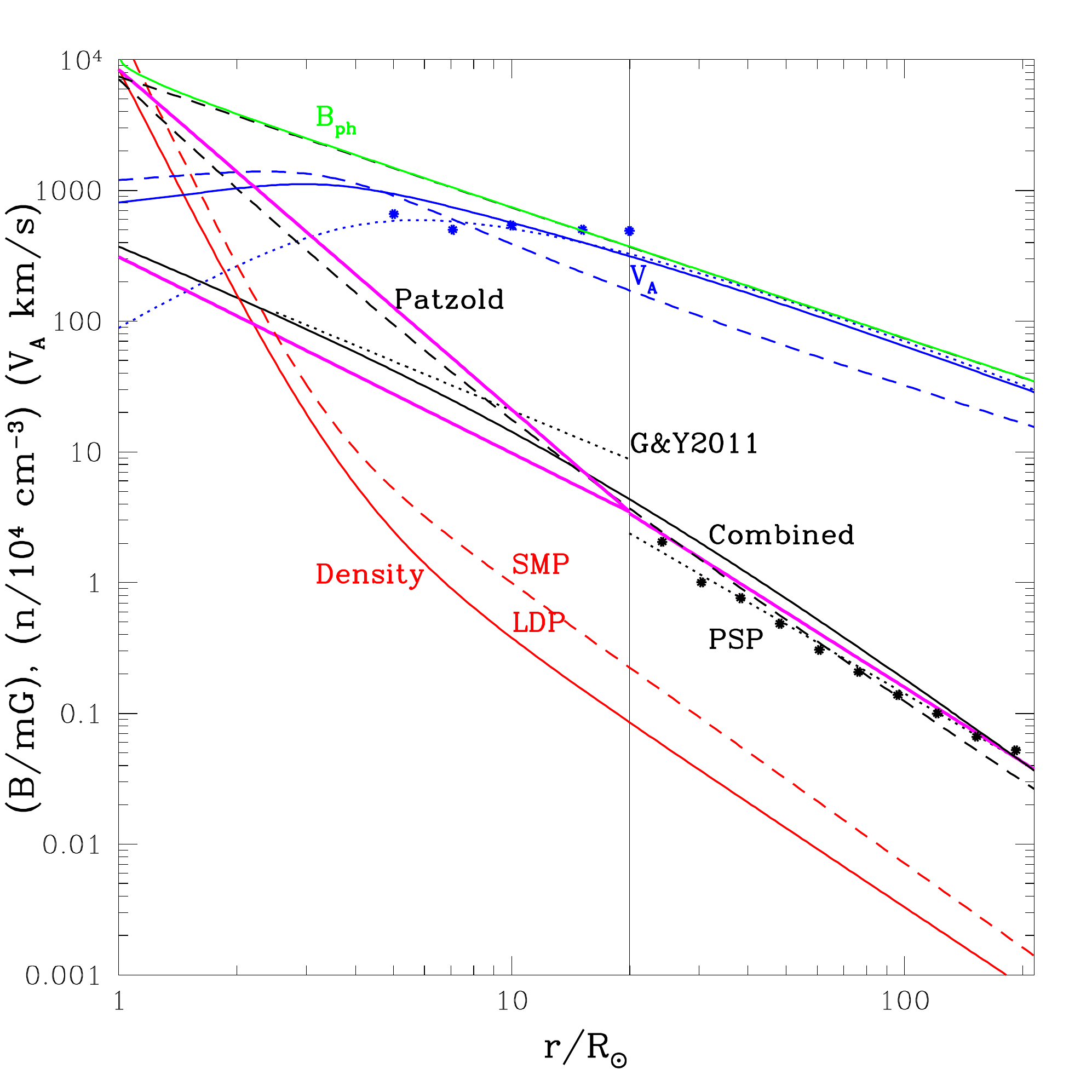}
			\includegraphics[width=3in]{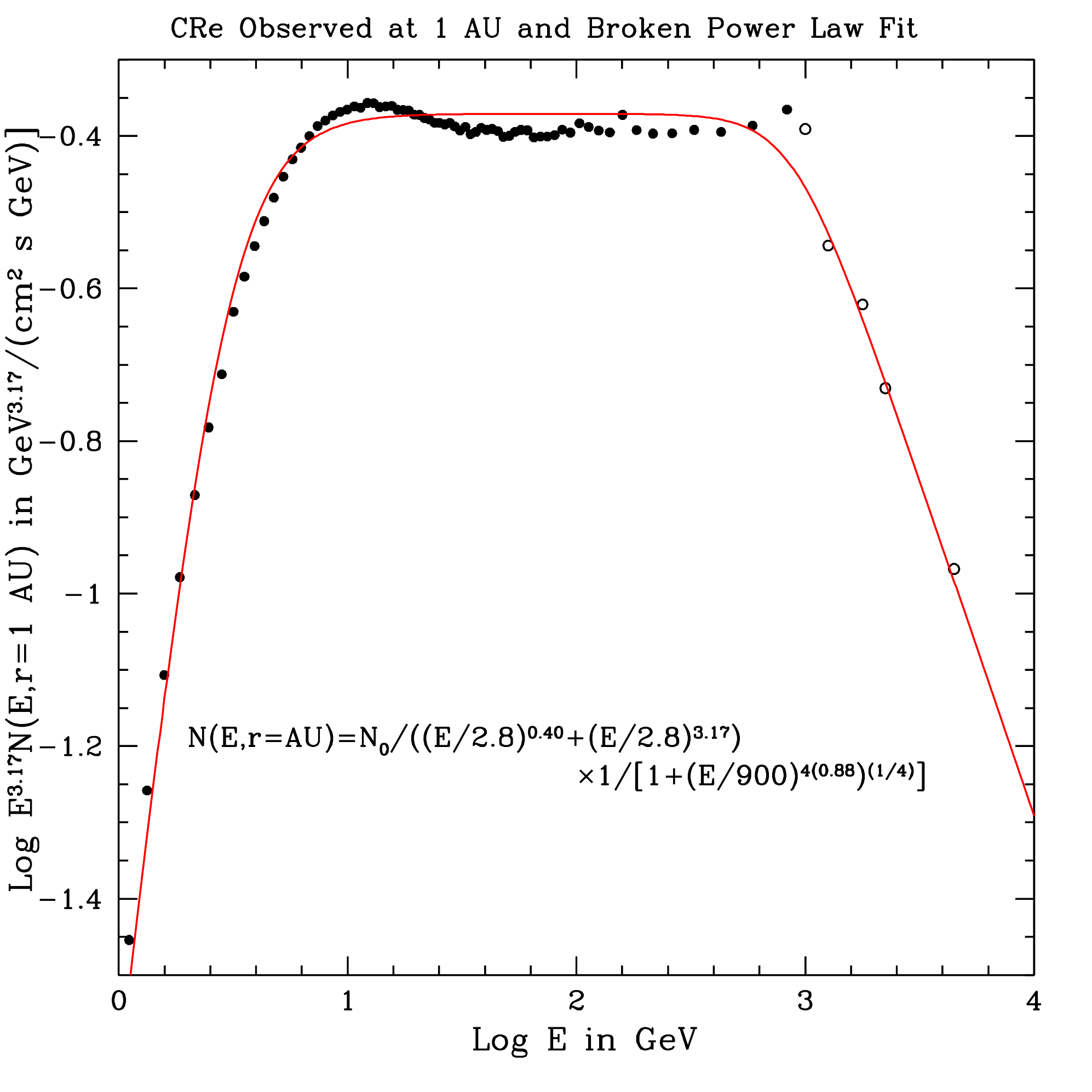}
			\caption{{\bf Left:} Some analytic fits to the observed structure of the magnetic field (black
				lines), density
				(red lines) and Alfv\'en velocity (blue lines). The black points are measurements from PSP (Badman et al. \,2021) The two dotted black lines show $B(r)$ fit to PSP points and to results from
				Gopalswamy \& Yashiro (2011; GY11).
				The solid line is a fit to combination of the two given
				in Eq. (\ref{CombinedB}). The dashed black curve is from Patzold et al.
				(1987). The upper solid green and dashed black curves shows the
				variation of  equivalent  field, $B_{\rm op}=\sqrt{8\pi u_{\rm op}}$, for two models of optical photons 
				energy density $u_{\rm op}$ given in Eqs. \ref{uphotonr} and \ref{uphotonfar}, respectively. The red curves show two models of the density from Saito et al. (1977,
				SMP) and Lehblanc et al. (1998, LDB). The blue curves present three models of the Alfv\'en velocity
				obtained using (i) the Combined+LDP (dotted), (ii) Patzold+SMP (dashed), and (iii) an average of the
				two (solid). The points are  Alf\'ven velocity values given by GY11. The three magenta lines show the
				three power law forms of Eq. (\ref{Bfits}) we use in our analysis. {\bf Right:} Observed spectrum of Cosmic Ray electrons by AMS02 (filled points) \citep{AMS02}  and H.E.S.S. \citep{HESS} (open circles)
				at 1 AU from the Sun. The red curve shows an analytic fit to the data consisting of a smoothly
				broken power-law with two breaks at energies $E_1= 2.8$ GeV (with index -0.4 below it) and $E_2=900$
				GeV (with index -4.05 above it), described by Equation (\ref{CReSpec}).
			}
			\label{bfieldetc}
		\end{center}
	\end{figure}

	Over the past decades there have been several models proposed for the variation of the magnetic
	field in the corona of the Sun and in the inner heliosphere ($r\leq 1$ AU). Also, there have been 
	several observations describing the $B(r)$ relation. In general there are large dispersion in the
	observed values but a power-law form, $B(r)\propto r^{-\d}$, provides a satisfactory fit. It is
	generally believed that the magnetic field in the heliosphere follows a Parker spiral with $\d\simeq
	2$. However, recent observations by the Parker Solar Probe (PSP) measurements at distances 
	$0.13<(r/{\rm AU})< 1$. or $27<(r/R_\odot) <214$,  shows some variation from this form with large dispersion, but on average they can be fit to a power law with $\d \sim 1.75$ and
	$B(r=1 {\rm AU})\sim 38 \mu$G (Badman et al. 2021). Gopalswamy \& Yashiro (2011; GY11) using
	observations
	of the Coronal Mass Ejections (CMEs) derive the variation of the $B$ field inside this
	region, $5<(r/R_\odot)< 25$ with $\d=1.27\pm 0.03$ and $B(r=5\rs)=0.05$ G.  These two nearly
	overlapping 
	observations,
	shown by the dotted lines on the left panel of Figure \ref{bfieldetc}, can be combined as 
	\beq
	\label{CombinedB}
	B(r)=0.4(r/\rs)^{-1.2}/(1+r/r_c), \,\,\, {\rm with}\,\,\, r_c=13\rs,
	\eeq
	shown by the solid black curve, which once extrapolated to  the photosphere yields $B_0\sim 0.4$ G.
	This  is smaller than
	$B_0\sim 10$ G indicated by lower corona  observations indicating that the profile must steepen
	rapidly below $5 \rs$ as indicated by other observations and models.
	For example, Patzold et al. (1987) give 
	\beq
	\label{BPatzold}
	B(r)=6(\rs/r)^3[1+r/(5\rs)] {\rm G}, 
	\eeq
	shown by the dashed black line that
	agrees with PSP observations and steepens to $B_0=7.2$ G at the photosphere.  Alissandrakis \& Gary
	(2021) describe some radio observations and present a summary of all past measurements.  There is a
	wide dispersion in these measurements as well. We will use a combination of these results in our treatment of transport and radiation of CRes.
	
	In what follows (for the $B$ field here and charcteristics of turbulence discussed below) we will
	treat
	the  outer region  ($0.1<r/{\rm AU}<1$) and the inner region ($1=r/\rs<20$) separately. For the
	outer region we use a fit to PSP observations and in the inner region we use the two widely
	different
	models, similar to GY11  and  Patzold  (1987) observations. These three fit forms, shown in magenta
	in Figure
	\ref{bfieldetc} (left), are
	
	\beq
	\label{Bfits}
	B(r)/{\rm G}=
	\begin{cases}
		1.0 (r/\rs)^{-1.9} & \, 0.1 < r/{\rm AU}  < 1\\
		0.31 (r/\rs)^{-1.5} & \, {\rm GY11}\\
		8.4 (r/\rs)^{-2.6} & \, {\rm Patzold}.
	\end{cases}
	\eeq
	GY11 also give two models of density variation, $n(r)$, due to Saito et al. (1977, SMP)
	and Lehblanc et al. (1998, LDB) shown on the left panel of Fig.~\ref{bfieldetc}
	by the dashed and solid red curves, respectively. The density and magnetic field variation allows us
	to calculate  the variation of  Alfv\'en velocity, $v_A=B/\sqrt{4\pi m_p n}$,  shown by the blue
	curves, which is needed for the treatment of  CRe transport described next. Here again we will
	use the following two approximate models;
	\beq
	\label{alfvenV}
	v_A = 
	\begin{cases}
		500(r/20\rs)^{-1.2}\, {\rm km\,s}^{-1}, & r>20\rs,\\
		v_A=500 \,{\rm km\, s}^{-1}, & r<20\rs.
	\end{cases}
	\eeq
	
	\subsection{Photon Energy Density Variation}
	\label{sec:uphoton}
	
	CRes will encounter photons radiated by the Sun that  have a black body frequency distribution with total flux
	\beq
	\label{BBflux}
	F_{bb}=\sigma_{\rm SB}T^4=L_\odot/(4\pi \rs^2),
	\eeq
	where $\sigma_{\rm SB}$ is the Stephan Boltzmann constant, $T$ is the surface temperature, and $L_\odot$ is the luminosity of the  Sun. In the optically thick ($\tau>1$) region just below the photosphere ($r\leq \rs$) the photon energy density $u_{\rm ph}(\rs^-)=4F_{bb}/c$ and in the optically thin region just above it the energy density of out-flowing photons will be half of this, $u_{\rm ph}(\rs^+)=2F_{bb}/c$. At larger distances where photons move radially the energy density approaches to 
	\beq
	\label{uphotonfar}
	u_{\rm ph}(r)=L_\odot/(4\pi r^2 c)=(F_{bb}/c)(\rs/r)^2 \,\,\,\, {\rm for}\,\,\,\, r>\rs.
	\eeq
	\citet{Orlando2007} derive the following relation describing the transition between the last two regions as:
	\beq
	\label{uphotonr}
	u_{\rm ph}(r)=2(F_{bb}/c)h(r) \,\,\, {\rm with} \,\,\, h(r)=1-\sqrt{1-(\rs/r)^2}.
	\eeq
	
	The photon energy density can be converted to an equivalent magnetic field, $B_{\rm ph}(r)=\sqrt{8\pi u_{\rm ph}(r)}$ with $B_{\rm ph}(\rs^+)=10.5$ G. The top (solid  green and dashed black) curves in Figure \ref{bfieldetc} shows  variations of $B_{\rm ph}(r)$ based on Equations (\ref{uphotonr}) and (\ref{uphotonfar}), respectively, which are significantly different at very small distances, $r<2\rs$.
	
	\subsection{The CR Electron Spectrum at 1 AU}
	\label{sec:crespec}
	
	The spectral intensity, $J(E, r=1{\rm AU})$, of the CRes at~1 AU  during the solar minimum
		are observed by AMS02 \citep{AMS02} and H.E.S.S. \citep{HESS}, which appears to be  highly isotropic. Thus, the total flux $F(E)=4\pi J$,
	which as
	evident from right panel of Figure \ref{bfieldetc}, obeys a power law with index $p=-3.17$, for the
	most relevant
	energy range of few GeV to TeV. For an analytic description, we fit the spectrum to a broken
	power law with two breaks at $E_1=2.8$ GeV with index $p_1=-0.40$ below it, and at $E_2=0.9$ TeV
	with index $p_2=-4.05$ above it:
	
	\beq
	\label{CReSpec}
	F(E,r=1 {\rm AU})=F_0\frac{(E/E_1)^{p_1}}{1+(E/E_1)^{p_1-p}} [1+(E/E_2)^{n(p-p_2)}]^{(-1/n)},
	\eeq
	where $F_0=0.014$ cm$^{-2}$ s$^{-1}$ GeV$^{-1}$, and $n=4$ for a sharper break. The total energy flux $F_{\rm tot}\sim 0.1$ GeV cm$^{-2}$ s$^{-1}$ is about $10^3$ smaller than solar wind energy flux.  The CRe spectral
	density $N(\g,r=AU)$ (needed for calculation of the emissivity) 
	is obtained by dividing the flux by the
	speed of light (for relativistic electrons) changing $E$ to $\g$ and $E_1, E_2$ to
	$\g_1=1957E_1=5.6\times 10^3,\g_2=1957E_2=1.8\times 10^6$. This gives $N_0=F_0/(1957c)=2.7\times 10^{-16}$ cm$^{-3}$  $\g^{-1}$, and total number density of $N_{\rm tot}\sim 3\times 10^{-12}$ cm$^{-3}$, again much smaller than the solar wind density.
	
	CRes with greater than GeV energy
	traveling  from 1 AU toward 
	the Sun will spiral around the magnetic field lines, initially with the
	above spectrum, and  an isotopic  pitch angle  distribution. During this
	transport they lose energy via the synchrotron and IC processes and are scattered by
	turbulence in the solar wind. Their pitch angle will also change due to these scatterings and the
	variation of  magnetic field. These interactions will  change their spectrum as described next.
	
	\section{Transport Effects and Spectral Variations}
	\label{sec:transport}
	
	\subsection{Transport Equation}
	\label{sec:transporteq}

	We first note that throughout inner heliosphere ($r<1$ AU) the  electron gyroradius,
	$r_g=v_\perp\g/(2\pi \nu_B)=1.7\times 10^3\g\sqrt{1-\mu^2}/B$ cm  is smaller than the size  of
	the source, or more precisely  the $B$ field scale height, 
	$H_B=(|{d{\rm ln} B/ds}|)^{-1}=1.2r/\d$. Here $v_\perp=v\sqrt{1-\mu^2}$ is the perpendicular component of the electron velocity and $\nu_B=eB/(2\pi m_ec)=2.8\times 10^6 B$ is the gyrofrequency, and for particles following Parker spirals   we use for the distance along  field lines $s\sim 1.2 r$.
	For isotropic distribution of pitch angles,  $\langle\sqrt{1-\mu^2}\rangle=\pi/4$, and for $B(r)=B_0(r/\rs)^{-\d}$ we have 
	\beq
	\label{gyror}
	\zeta_g\equiv r_g/H_B=1.6\times 10^{-8}\g\d(r/\rs)^{\d-1}\times({\rm G}/B_0).
	\eeq
	In the most relevant inner region, for the Pazold model with $B_0\sim8.4$G, $\d\sim 2.6$, this ratio is  $\zeta_g=0.6(\g/10^6)$ and $=0.002(\g/10^6)$ at $r=20\rs$ and 1, respectively. For the GY11 model these ratios are $\zeta_g=0.4(\g/10^6)$ and $\zeta_g=0.008(\g/10^6)$, indicating that throughout most of the inner region CRes are tied to the magnetic field lines and spiral down to the Sun along  Parker spiral guide fields. Thus, inside this radius  the  modulation approach is not appropriate and we need a kinetic approach. On the other hand, in the outer region  with $B_0=1$ G and $\d\sim 2$, $\zeta_g=4(\g/10^6)$ and $\zeta_g=0.5(\g/10^6)$ at 1 AU and $r=20\rs$, respectively, so that  the kinetic approach provides an approximate description of the transport. However, since most of energy loss and emission occurs mainly near the Sun this approximation would be adequate.%
	\footnote{Note that this will also be the case for protons with $\g<10^3$ or energies less than 1 TeV, as is the case for electrons.}
	This also implies that in the region with $\zeta_g<1$  we can use the gyro-phase averaged particle density distribution
	$f(t, s, \mu, E)$ as a function of time, distance, $s$, pitch angle cosine, $\mu$,
	and energy, $E$ (and velocity $v$).
	
	This distribution  can be described by the following version of the Fokker-Planck transport equation.
	
	\beq
	\label{KE}
	\frac{\partial f}{\partial t}+v\mu\frac{\partial f}{\partial s}  - \frac{v \partial {\rm ln} B}{2
		\partial s}\frac{\partial}{\partial\mu}\left[(1-\mu^2)f\right]-\frac{\partial}{ \partial \mu}
	\left[D_{\mu\mu} \frac{\partial f}{\partial \mu}\right]=\frac{\partial ({\dot E}f)} {\partial E}+{\dot
		Q(t,\mu,E)},  
	\eeq
	where ${\dot E}$ is the absolute value of the energy loss rate,
	$D_{\mu\mu}$ is the pitch angle diffusion rate,%
	\footnote{We ignore energy diffusion rates, $D_{EE}$, which for relativistic particles is $(v_A/c)^2<10^{-5}$ times smaller than $D_{\mu\mu}$ throughout the heliosphere. We also ignore terms involving solar wind, Alfv\'en and other drift velocities, which are much smaller than the CRe speed, $v=c$.}  
	and ${\dot Q}$ describes the energy spectrum and pitch angle distribution  of the injected particles at  1~AU, $s=0$. In what follows, instead of $s$, we use distance from the Sun, $r=1~AU-s/(1.2)$  or $ds\sim 1.2 dr$, which implies we multiply $D_{\mu\mu}, {\dot E}$ and ${\dot Q}$ by 1.2. Since all coefficients of this equation ($B, D_{\mu\mu}, {\dot E}, {\dot Q}$) vary on time scales much longer than the transport time of CRes from 1 AU to the Sun, we can assume steady state, i.e.~we can set $\partial f/\partial t=0$, and set the  injection rate at 1 AU to ${\dot Q}(E)= F(E, r=AU)$, given by Equation (\ref{CReSpec}), with an isotropic pitch angle distribution. Then the spectral flux down to the Sun will be $F(r, E)=v\langle \mu\rangle f(r, E)=vf(r,E)/2$. 
	
	It should also be noted that the above analysis is valid when diffusion of particles perpendicular to the magnetic field is small compared to diffusion parallel to the field, described by $D_{\mu\mu}$. Approximately, this requires a particle gyro-radius less than its mean free path described in \S \ref{sec:scat}. As shown in Appendix C, this is satisfied for $\g\leq 10^6$ throughout most of the inner heliosphere, especially in the inner regions near the Sun where  losses are most important.  
	
	We note that the kinetic approach is very different than the common use of modulation potential, which seems to work well in the outer ($r> 1$ AU) heliosphere, but its extrapolation  to the inner regions is highly uncertain. As will be shown below, we obtain different spectral variation with the kinetic approach.
	
	We now give  detailed description of the transport coefficients.
	
	\subsubsection{Energy Loss}
	\label{sec:eloss}
	Throughout most of the outer heliosphere the {\bf energy loss}, described by the first term on the right hand side of Equation (\ref{KE}), is negligible but it increases
	relatively rapidly with energy, and as the CRes approach the Sun. Relativistic electrons with isotropic pitch angle distribution lose energy
	mainly by IC and synchrotron processes%
	\footnote{Bremsstrahlung losses may become important below the photosphere, which will not be of
		interest here. Bremsstrahlung may also be more important than synchrotron at distances $\geq 1$ AU and low energies
		where all losses are negligible.}
	with the rate (see, e.g. Eqs. 7.16 and 7.17 of \citet{RL})
	\beq
	\label{eloss}
	{\dot E}=(4/3)c\sigma_T(\g^2-1)u_{\rm eff} \,\,\, {\rm with} \,\,\, u_{\rm eff}= u_{\rm ph}+B^2/(8\pi).
	\eeq
	where  $\sigma_T=6.6 \times 10^{-25}$ cm$^2$. Replacing $u_{\rm ph}=B^2_{\rm ph}(\rs^+)/8\pi$ and defining $B_{\rm eff}^2=B^2_{\rm ph}(\rs^+)+B^2$ we obtain the rate of change of  the Lorentz factor, $\g=E/(m_ec^2)$,
	
	\beq
	\label{dgammdt}
	d\g/dt=1.27\times 10^{-9}(\g^2-1)(B_{\rm eff}/{\rm G})^2 \, {\rm s}^{-1}.
	\eeq
	In section \ref{sec:CReandRadFlux} we will need variation of $\g$ with distance $d\g/dr=1.2 d\g/ds=[1.2/(\langle \mu\rangle c)]d\g/dt$.
	In what follows we use the dimensionless distance $x=r/\rs$ so that for $B(r)=B_0x^{-\d}$ we can write $B^2_{\rm
		eff}=[B_{\rm ph}(\rs^+)]^2[h(x)+\zeta x^{-2\d}]$, where $\zeta=[B_0/B_{\rm ph}(\rs^+)]^2=(B_0/10.5 {\rm G})^2$. Thus, we  obtain 
	
	\beq
	\label{eloss1}
	d\g(r)/dx=7.8\times 10^{-7}(\g^2 -1)(h(x)+ \zeta x^{-2\d}). 
	\eeq
	We note the following three important aspects of the above loss rate. 1. For $B_0\sim 10$ IC and synchrotron losses will be comparable near the Sun, but since $\d\sim 2$, IC losses will dominate at larger distances. However, since most of the radiation is produced near the Sun the IC energy emission in gamma-rays and synchrotron in UV-X-ray range will be comparable. 2. For free streaming relativistic electrons near the Sun ($x\sim 1$), 
	$\D\g/\g \sim 1 $ for TeV electrons, so that energy losses cannot be ignored. In addition, as shown
	below the free streaming
	assumption is not correct and particles take longer time to reach the Sun and hence lose more
	energy. 3. The IC loss rate ignores the Klein-Nishina (K-N) effect, which reduces the rate at $\g> \g_{\rm KN}\sim m_ec^2/\epsilon_{\rm ph}\sim 5\times 10^5$.  For photon energy $\epsilon_{\rm ph}\sim 1$ eV, approximately by a factor $f_{\rm KN}\sim 1/(1+t^2)$ or $1/(1+t)^{3/2})$ according to \citet{Hooper} and \citet{mod05}, respectively with $t=\g/\g_{\rm KN}$, indicating than K-N effect can be ignored for electron energies below few 100 GeV. Numerical calculations  by \citet{OrlandoThesis} shows that at $f_{\rm KN}\sim 0.5$ for TeV electrons. We will ignore K-N effects in this preliminary analysis of the transport.
	
	\subsubsection{Advection and Crossing Time}
	\label{sec:taucross}
	
	The second term in Equation (\ref{KE}) describes particle advection and can be characterised by the crossing time across a source of size $L$ as $\tcross\sim L/v$. In our case we set $L=s=1.2 r$, and $v=c$ to obtain
	\beq
	\label{tcross}
	\tcross=2.8x\,\,{\rm s}. 
	\eeq
	
	\subsubsection{Pitch Angle Variations}
	\label{sec:scat}
	
	The final important aspect of transport involves pitch angle changes which are caused by the following three processes.
	
	\begin{enumerate}
		
		\item 
		
		Pitch angles change due to the {\it energy loss processes} of relativistic electrons is negligible. For
		example,
		for the synchrotron process 
		\beq
		\label{muchange}
		\frac{d\mu}{dt}= \left(\frac{c\sigma_T}{4\pi m_ec^2}\right)\mu(1-\mu^2)B^2/\g \,\,{\rm cm}^{-1},
		\eeq
		which is $\g^{-3}$ times smaller than the energy loss rate, $\frac{d\gamma}{dt}$  (see, e.g.~Petrosian 1985) and can be ignored in Equation (\ref{KE}). 
		
		\item
		
		The pitch angle will also  change  because of the
		{\it convergence of the $B$ field}  by the large factor of $>10^4$ during transport from  1 AU to the Sun. This effect is described by the third term in Equation (\ref{KE}), with the characteristic time scale, $\tau_B=2H_B/v$, where the magnetic field scale height is $H_B=(|
		\frac{\partial {\rm ln} B}{\partial
			s}|)^{-1}$. For a power law $B(r)$ with index $\d$, $ds=1.2 dr$, and $v=c$ this  yields the time scale 
		\beq
		\label{tauB}
		\tau_B=5.6 x/\d \,\,{\rm s}.
		\eeq
		In the presence of such strong 
		convergence  only electrons within a narrow 
		pitch angle range (those in the loss cone)  can reach the Sun, thus, requiring an efficient
		scattering process to scatter the particles into the loss cone that will allow  transport to the
		Sun. 
		
		\item
		
		The third and most important cause of pitch angle change is {\it scattering by turbulence}.
		
		As is well known, the solar
		wind, through witch the CRes propagate,  contains high level of turbulence, which can be the
		scattering agent. This process is governed by the pitch angle diffusion coefficient with the characteristic scattering time  or mean free path of (see Petrosian 2012 and the discussion below)
		\beq
		\label{tscat}
		\tsc=(3/8)\int_{-1}^1\frac{(1-\mu^2)^2}{D_{\mu\mu}}d\mu  \,\,\, {\rm and} \,\,\, \l_{\rm mfp}=v\tsc.
		\eeq
		As will be shown below this time for most part is smaller than the above two transport time scales  (Eqs. \ref{tcross} and \ref{tauB}).
	\end{enumerate}
	
	\subsubsection{Escape Time}
	\label{sec:tesc}
	
	The combined effect
	of these  processes  determines the resident or travel time of particles at any point,  and the time for traverse from 1 AU to the Sun, denote by an escape time $\tesc(r)$, which is a function of the above defined time scales.
	
	The exact treatment of this problem requires numerical solution of the Fokker-Planck kinetic equation. This is beyond the
	scope of the current paper. Here we use some approximate treatment based on some analytic results
	from Malyshkin \& Kulsrud (2001) and numerical simulations of Effenberger \& Petrosian (2018). As shown in these papers,  in the {\bf strong diffusion limit}, i.e.~when $\l_{\rm
		mfp}
	\ll r\, {\rm and} \, H_B$, or equivalently when $\tsc \ll (\tcross \, {\rm and} \, \tau_B)$, the pitch angle distribution will remain
	isotropic. First consequence of this is that the
	important factor is not the large total convergence factor but the local convergence factor $\eta=r/H_B\sim \d/1.2<2$, meaning the presence of a relatively large loss cone.
	Second, in this case we are dealing with the well known random walk relation 
	and $\tesc\sim \tcross^2/\tsc\gg \tcross$. More generally, in this case  one can use pitch-angle integrated (in the downward directions) quantities in Equation (\ref{KE}). Defining total density $N(s, E)=\int_{-1}^1 d\mu\, f(t,s,\mu,E)$ and ${\dot Q}(t, E))=\int_0^1d\mu \, {\dot Q}(t,\mu, E)$, we obtain
	\beq
	\label{FPiso}
	\frac{\partial N}{\partial t}= \frac{\partial}{\partial s}\kappa_{ss}\frac{\partial N}{\partial s} 
	+ \frac{\partial (\langle\dot{E}\rangle N)}{\partial E}  + {\dot Q}(t,E) \,\,\,\, {\rm
		with}\,\,\,\,\kappa_{ss} = \frac{v^2}{8}\int_{-1}^1 d\mu\frac{(1-\mu^2)^2}{D_{\mu\mu}}.
	\eeq
	where $\kappa_{ss}$ is the spatial diffusion coefficient related to the scattering time (or mean free
	path) as $\tsc=3\kappa_{ss}/v^2$. We show in Appendix D, that if one ignores the energy loss term, which is the common practice, this equation can be solved approximately, yielding  spectral variation  that is quantitatively different than the one derived using the modulation approach.
	
	There are no simple analytic solutions when  the energy loss term is included. However, simple dimensional argument, or spatially integrated version of this equation, known as the leaky box model, implies that, in the strong diffusion limit, $\tsc\gg \tcross$, time spent in a region of
	size $L$, or the escape time $\tesc\sim L^2/\kappa_{ss}\sim (\tcross)^2/\tsc$, as deduced from the random walk problem.
	In the opposite weak diffusion limit ($\tsc\ll \tcross\sim \tau_B$), the
	particles are reflected and can escape toward the Sun only when scattered into the loss cone. Thus, the escape time becomes proportional to the scattering time, and as shown in the above papers the
	proportionality constant is equal to the logarithm of  convergence factor $\eta=\d/1.2$. The numerical simulations of Effenberger \& Petrosian (2018),
	based on the Fokker-Planck equation, show that the following relation, similar to Malyshkin \& Kulsrud  (2001)
	equation, provides an excellent approximation for the isotropic pitch angle scattering rate and
	isotropic distribution of the injected particles:
	
	\beq
	\label{MKEP}
	R(r,\g)=\tesc/\tcross = \tcross/\tsc + 2\eta + \ln{\eta}(\tsc/\tcross)\,\,\,\, {\rm with}\,\,\,\, \eta\sim
	\d/1.2. 
	\eeq
	The first and third terms on the right hand side describe the above two limiting cases that are
	connected by 
	the middle term with $R$ nearly constant (independent of scattering time) and close to  the minimum 
	value, $R_{\rm min}=2(\eta + \sqrt{\ln \eta})$ (at $\tsc=\tcross/\sqrt{\ln \eta}$), which varies from
	$\sim 3$ to 6 for $\d=1.5$ to 2.6. 
	
	This equation involves the three time scales defined above,  which vary with distance from the Sun, with the critical variable being the mean free path, or the scattering time, $\tsc$, that depends also on particle energy. As described below it depends on the energy density and spectrum of the turbulence, in addition to magnetic field and gas density in the solar wind.
	
	This procedure allow to separate the implicit dependence on distance of $N(s, E)$ (described in Appendix D) and energy dependence described next.  The upshot of this is that the transport time, $\tesc$, of CRes is longer than free 
	crossing time, $\tcross$, which, as stated above, increases the energy losses by the factor $R(r,\g)$, and directly affects spatial dependence.
	
	\subsection{Energy Loss Enhancement Factor}
	\label{sec:RofG}
	
	The Energy loss enhancement factor (ELEF)  depends on the three time scales defined in previous section, $\tcross,\tau_B, \tsc.$ The first two are well defined given $B(r)$. 
	Evaluation of the third is more complicated because there are no direct measurements of $\l_{\rm mfp}$ or  $\tsc$, so  we have to rely on theoretical models of turbulence-particle interaction rates, which, in addition to $B(r)$  and background particle
	density, $n(r)$, requires measurement of  the characteristics   of turbulence and its variation with distance. 
	
	\subsubsection{Characteristics of Turbulence}
	\label{sec:turbulence}
	
	Over the last two decades there have been several in situ measurements of the
	intensity and spectrum of turbulence, ${\cal W} (k)$,  around 1 AU by near-Earth instruments (see, e.g.~Leamon et al.
	1998, 1999; Bruno and Cordone, 2013).  More recently observations by PSP (Chen et al. 2020) have
	extended this information from 1 to $\sim 0.1$  AU. In Appendix A we summarize  these
	measurements, ending with a parametrized form for the  turbulence energy density and its variation with distance from the Sun, ${\cal W} (r, k)={\cal W}_{\rm turb}(r) k^{-q}$, where $q$ is the spectral index in the inertial range, $k_{\rm min}<k<k_{\rm max}$. Measurements around 1 AU indicate that $q=5/3$ (Kolmogorov) but PSP measurements indicate a gradual variation from $q=5/3$ to Iroshnikov-Kraichnin (I-K) index of $q=3/2$ between 0.3 to 0.2 AU. In our analysis we will use both these models plus a model with $q=2$ that stands for the free transport case (i.e.~unhindered by the $B$ field variation or turbulence). We fit the observed spatial variation to a power law for the outer region ($r\geq 20\rs$). The result, shown by the solid blue line in Figure \ref{fig:TurbParams}, is 
	\beq
	\label{turb}
	{\cal W}_{\rm turb}(r)={\cal W}_0(r/\rs)^{-\d_{tr}}, \,\,\, {\rm with}\,\,\, \d_{tr}=3.1,\,\,\,{\cal
		W}_0=0.033\,\, {\rm nT}^2.
	\eeq
	
	For the inner region we use a combination of extrapolation of the above expression and some theoretical results. In Appendix A we also present expression for the ratio of turbulence to magnetic energy  densities, $f_{\rm turb}={\cal W}_{\rm turb}(r)/B^2(r)$ needed below.
	
	\subsubsection{The Scattering Time}
	\label{sec:tscat}
	
	Theoretical models of wave-particle interaction rates determine  the scattering time, which depends in a complicated way on several variables and parameters  related to $B(r), n(r)$ and ${\cal W} (r, k)$. As shown in Appendix B there are two main parameters. The first is the ratio of plasma to gyro-frequencies, $\alpha\equiv \omega_p/\Omega\propto \sqrt{n}/B$, a measure of the degree of magnetization or the Alfv\'en velocity in units of the speed of light, $\b_A=v_A/c$; for protons $\a_p=1/\b_A$, for electrons $\a_e=\sqrt{(m_e/m_p)}/\b_A$. The second is the characteristic wave-particle  time scale $\tau_p$ or the rate (see, e.g. Dung \& Petrosian, 1994)
	\beq
	\label{taup}
	\tau_p^{-1}=(\pi/2)\Omega f_{\rm turb}(q-1)\Phi^{(q-1)} \,\,\,\, {\rm with}\,\,\,\, \Phi=ck_{\rm
		min}/\Omega=f_{\rm min}/(\b_A\nu_B),
	\eeq
	with electrons gyrofrequency $\nu_B=\Omega_e/(2\pi)=2.8\times 10^6$ Hz.
	At low energies and high magnetization ($\a_e<1$) electrons interact with many plasma waves complicating the results (Pryadko \& Petrosian 1997, 1999; Petrosian \& Liu 2004). However, for relativistic electrons, with Lorentz factor $\g>m_p/m_e$, and low magnetization (i.e.~$\a_e\gg 1$ or $\b_A\ll 1$, which is the case in the solar wind), electrons interact only with low frequency (or small $k$) Alfv\'en waves, with the dispersion relation $\omega(k)=v_Ak_\|$, for parallel propagating waves,  and with fast mode waves, $\omega(k)=v_Ak$, for perpendicular propagating waves, both kind of which are present in the solar wind. In this case the energy dependence of scattering time simplifies to  $\tsc(r, \g)=\tsc_{,0}(r)\g^{2-q}$, where, as shown in Equation (\ref{tscat2}),  $\tsc_{,0}(r)/\tau_p=1.6, 2.6$, and 3.9 for $q=3/2, 5/3$, and 2, respectively. Using the observed characteristics of the turbulence in the outer region and its extrapolation to the Sun, and the three models of $B$ field, we can calculate the ratio $\tcross/\tsc$ and the ELEF $R(r,\g)=\tesc/\tcross$. As shown below, in most part we are in the strong diffusion limit so that $R(r,\g)=\tcross/\tsc$. 
	
	The radial variations of $R(r,\g=1)= \tcross/\tsc_{,0}$ (modulo the value of $\xi$ defined in Appendix B) are shown in the left panel of Figure \ref{fig:Rofx}. As evident, the extrapolation of the outer region curves to the Sun lies roughly half way between the widely different inner curves due to difference in the $B$ filed models there. For example, this ratio is 50(600) for $q$= 5/3(3/2) at the Sun. 
	These values, and the  radial variations are not
	too dissimilar to the theoretical estimation of the mean free path by Fichtner et al. (2012), shown by the dotted line.  As evident for most part this ratio is greater than 1 so we are in the {\bf strong diffusion limit} and  the ELEF can be approximated as 
	
	\beq
	\label{Rform}
	R(r, \g)=\tcross/\tsc=R_0(r/\rs)^{\epsilon}\g^{(q-2)}.
	\eeq
	
	From the description of  derivation of this ratio given in Appendix B it is easy to show that
	$R\propto \tcross{\cal W}_0(r)k_{\rm min}^{(q-1)}/B^q$ so that (for $\tcross \propto r$ and $k_{\rm min}\propto 1/r$) $\epsilon=2-\d_{tr}+q(\d-1)$, which gives  $\epsilon=0.25, 0.40  \, {\rm and} \, 0.70$ for $q=3/2, 5/3 \, {\rm and} \, 2$, respectively. With this extrapolation to $r=\rs$, and setting $\xi=0.04$ we obtain $R_0=3000, 200$ and 1, respectively. However, for $q=2$ we will use $\epsilon=0.0$  as  a proxy for the free transport case ignoring scattering and field convergence effects. We note that ELEF decreases with energy
	and the assumption of strong diffusion limit will not be valid at energies $\g>\g_{\rm max}=(R_0x^\epsilon)^{1/(2-q)}$. However, even at the photosphere  $\g_{\rm max}\sim 4\times 10^6 \, {\rm and} \, \sim 2\times 10^6$ for $q=3/2$ and $5/3$, respectively. For $q=2$ the ELEF is independent of energy but because of field convergence effect  we may be in the middle region of equation (\ref{MKEP}), with $R_0=R_{\rm min} \sim 3$, a value larger than 1, so our value of $R_0=1$ for $q=2$ gives the absolute minimum effect of the energy loss on the spectrum.
	
	It should be noted that the above values of ELEF parameters are uncertain because of absence of measurements of turbulence characteristics at $r< 20\rs$. Thus, the results presented below based on these parameters  should be considered as a representative of range of the possible effects of the energy loss. Our main goal, to be dealt with in upcoming papers, is to use  \f and \r observation for constraining these parameters.

	\begin{figure}[!ht]
		\begin{center}
			\includegraphics[width=3in]{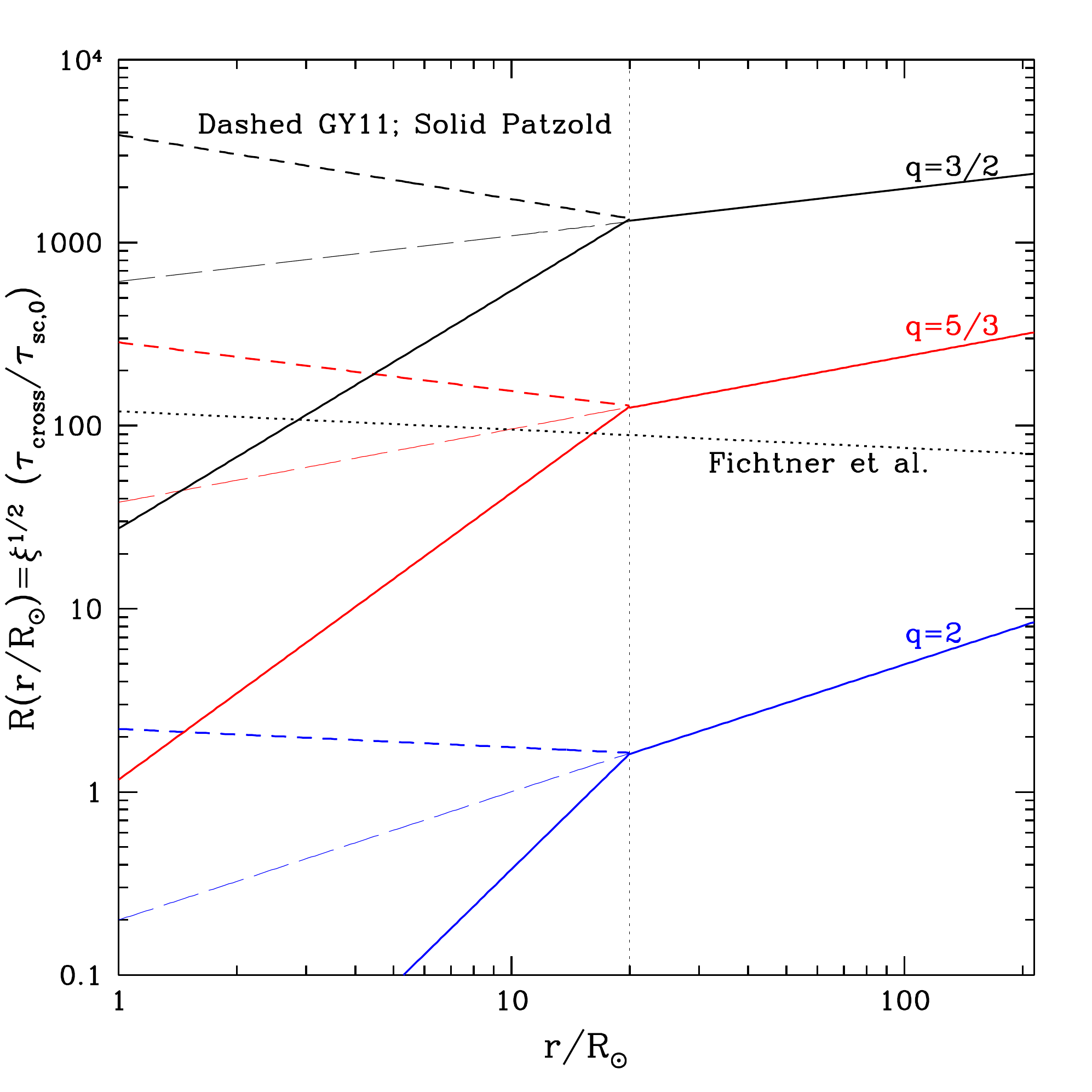}
			\includegraphics[width=3in]{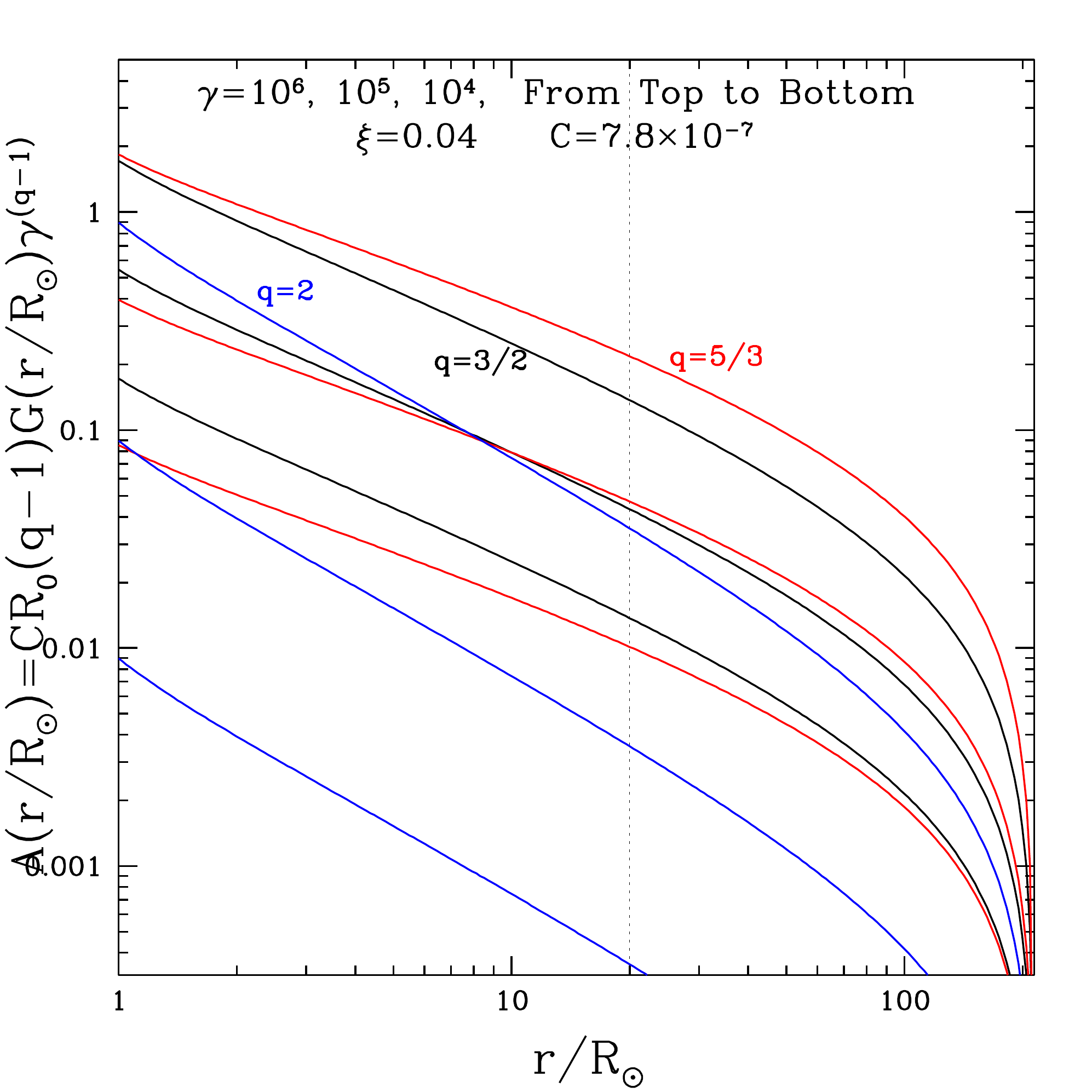}
			\caption{{\bf Left:} Spatial variation of the ratio $\tcross/\tsc_{,0}=R(r/\rs, \g=1)$ (modulo the scaling factor
				$\xi^{1/2}$) for I-K model with $q=3/2$,  Kolmogorov with $q=5/3$, and free transport approximation $q=2$. Note that the ratio $R(r, \g=1$ is very different for the three models at $\g=1$ but because of different energy dependence of the models the differences  at the relevant energies $10^4<\g<10^6$ is much smaller.  {\bf Right:} Radial variations of
				$A(r/\rs, \g)$ defined in Equation (\ref{Aofx}), for three values of $\g$ and  three values of $q$; $q=3/2, \epsilon=0.25, R_0=3000$, $q=5/3, \epsilon=0.40, R_0=200$, and $q=2, \epsilon=0, R_0=1.0$. As evident the differences between the models at high energies is smaller.}
			\label{fig:Rofx}
		\end{center}
	\end{figure}
	
	\section{CRe Spectral Variation and Resultant Radiation Flux}
	\label{sec:CReandRadFlux}
	
	As described above, CRes with the spectrum observed at 1 AU are guided along $B$ fields following Parker spirals)  to the Sun maintaining the isotropic pitch angle distribution because of scattering by turbulence. This process  changes (1) their energy loss rate and thus their spectra, and (2) their number density because of the reduction of their bulk flow and focusing by converging magnetic fields.

	\subsection{Energy Loss Rate and CRe Spectral Variations} 
	\label{sec:specvar}
	
	The  energy loss rate given  by Equation (\ref{eloss})  
	is enhanced by the factor $R(r,\g)$. Thus,  
	multiplying the energy loss rate given in Equation (\ref{eloss1}) by $R(\g, r)$ in Equation (\ref{Rform}) we obtain
	\beq
	\label{eloss2}
	d\g/\g^q=7.8\times 10^{-7}R_0g(r/\rs)d(r/\rs),\,\,\,\, g(x)=x^\epsilon(h(x)+\zeta x^{-2\d})
	\eeq
	Integrating this for an electron with initial Lorentz factor $\g_0$ at 1 AU ($x_{au}=214$) gives the
	variation of Lorentz factor with distance as 
	
	\beq
	\label{gammar}
	\g^{(1-q)}(r)=\g_0^{(1-q)}+CG(r/\rs) \,\,\,\, {\rm with} \,\,\,\, G(x)=\int_x^{x_{au}} g(x)dx; \,\,\,
	C=7.8\times 10^{-7}(q-1)R_0.
	\eeq
	There is no simple analytic expression for $G(x)$. For this purpose we set $h(x)=0.5(x^{-2}+x^{-n})$, with $n\gg 1$ to account for the sharp increase of $h(x)$ as $x\rightarrow 1$. Using this form, which gives identical value for $G(1)$ for $n=8$, we obtain 
	\beq
	\label{Gofx}
	G(x)=0.5[H_{(1-\epsilon)} + H_{(n-1-\epsilon)}] + \zeta H_{(2\d-1-\epsilon)} \,\,\, {\rm with} \,\,\, H_a=(x^{-a}-x_{au}^{-a})/a.
	\eeq
	From this we obtain
	\beq
	\label{gammazero}
	\g_0=\left(\frac{\g}{[1-A(x, \g)]^{1/(q-1)}}\right) \,\,\,\, {\rm and} \,\,\,\,
	\frac{d\g_0}{d\g}=\left(\frac{\g_0}{\g}\right)^q,
	\eeq
	where we have defined the critical function
	\beq
	\label{Aofx}
	A(x, \g)= CG(x)\g^{(q-1)},
	\eeq
	which is shown in Figure \ref{fig:Rofx}
	(right) for $\xi=0.04$,  three values of $q$,  and Lorentz factors $\g=10^6, 10^5$ and $10^4$.
	As evident this crucial factor has  similar distance dependencies for the three models.
	However,  the dependence on energy (or the Lorentz factor) is more variable. 
	
	Given $A(x, \g)$   we obtain the spectral variation with $r$  of CRe density%
	\footnote{We use number density rather than flux since  in calculating emissivity in Eq. (\ref{emissivity}) we need number density.} 
	as $N(\g, r)=
	N(\g_0, r)d\g_0/d\g=N(\g_0, r)(\g_0/\g)^{q}$.  Setting the energy dependence of $N(\g_0, r)$ to  the observed spectrum at 1 AU
	given in Equation (\ref{CReSpec}), (and changing $E$ to $\g$) we obtain the spectral shape at different distances inside 1
	AU, due to energy loss  as%
	\footnote{Note that we require $A(x,\g)>1$, which means the spectra at small distances $\rightarrow 0$ for $\g=\g_{\rm max}\equiv[CG(x)]^{1/(q-1)}$.}
	\beq
	\label{Nofr}
	N(\g, r)=N_0(r)\left(\frac{\g}{
		[1-A(r/\rs)]^{1/(q-1)}}\right)[1-A(r/\rs)]^{-q/(q-1)},
	\eeq
	where $N_0(r)$ describes the spatial variation, with $N_0(r=1 {\rm AU})=N_0$ defined below Equation (\ref{CReSpec}). The (correction) term in the square brackets will be more important  at higher 
	energies and  closer to the Sun with maximum value at the
	photosphere with \,$G_{\rm max}=G(1)= 0.5/(1-\epsilon) +0.5/(n-1-\epsilon) + \zeta/(2\d-1-\epsilon)\sim 1$.
	
	\subsection{Spatial Variations}
	\label{sec:spacialvar}
	
	As mentioned above the spatial variation is affected by two processes. First, in the regions where gyro-radius is smaller than $B$ field scale height $H_B$ (i.e.~$r<r_{\rm cr}$), the convergence  of field lines toward the Sun focuses the particles, so that their number density  in a bundle of field lines increases inversely with cross sectional area, $A(r)$, of the bundle and  $N_0(r)\propto 1/A(r)\propto r^{-2}$ for radial or Parker spiral field lines. Second, interactions with turbulence changes the CRe residence time, or the escape time, $\tesc$.  This changes the normalization by  the ratio $\tesc/\tcross$, which in the strong diffusion limit is equal to $\tcross/\tsc_{,0}=R_0(r)=R_0(r/\rs)^\epsilon$. The combine effect then yields
	\beq
	\label{N0ofr}
	N_0(r)=N_0(r_{\rm cr})\left(\frac{A(r_{\rm cr})R_0(r)}{ A(r)R_0(r_{\rm cr}))}\right)=N_0(r_{\rm cr})(r/r_{\rm cr})^{\epsilon-2}.
	\eeq
	As shown in Appendix D, this spatial variation can be derived by integration of Equation (\ref{FPiso}) over the volume of a bundle of field lines.
	
	\subsection{CRe Spectarl Variation}
	\label{sec:CReSpecVar}
	
	Because of the uncertainty in  value of $r_{\rm cr}$, here we focus on effects of commonly ignored  energy loss which gives the dependence of flux on energy setting $N_0(r)=N_0$. Inclusion of the spatial variation in Equation (\ref{N0ofr}) will scale the energy spectra by  $(r/r_{\rm cr})^{\epsilon-2}$.
	
\begin{figure}[!ht]
		\begin{center}
			\includegraphics[width=2in]{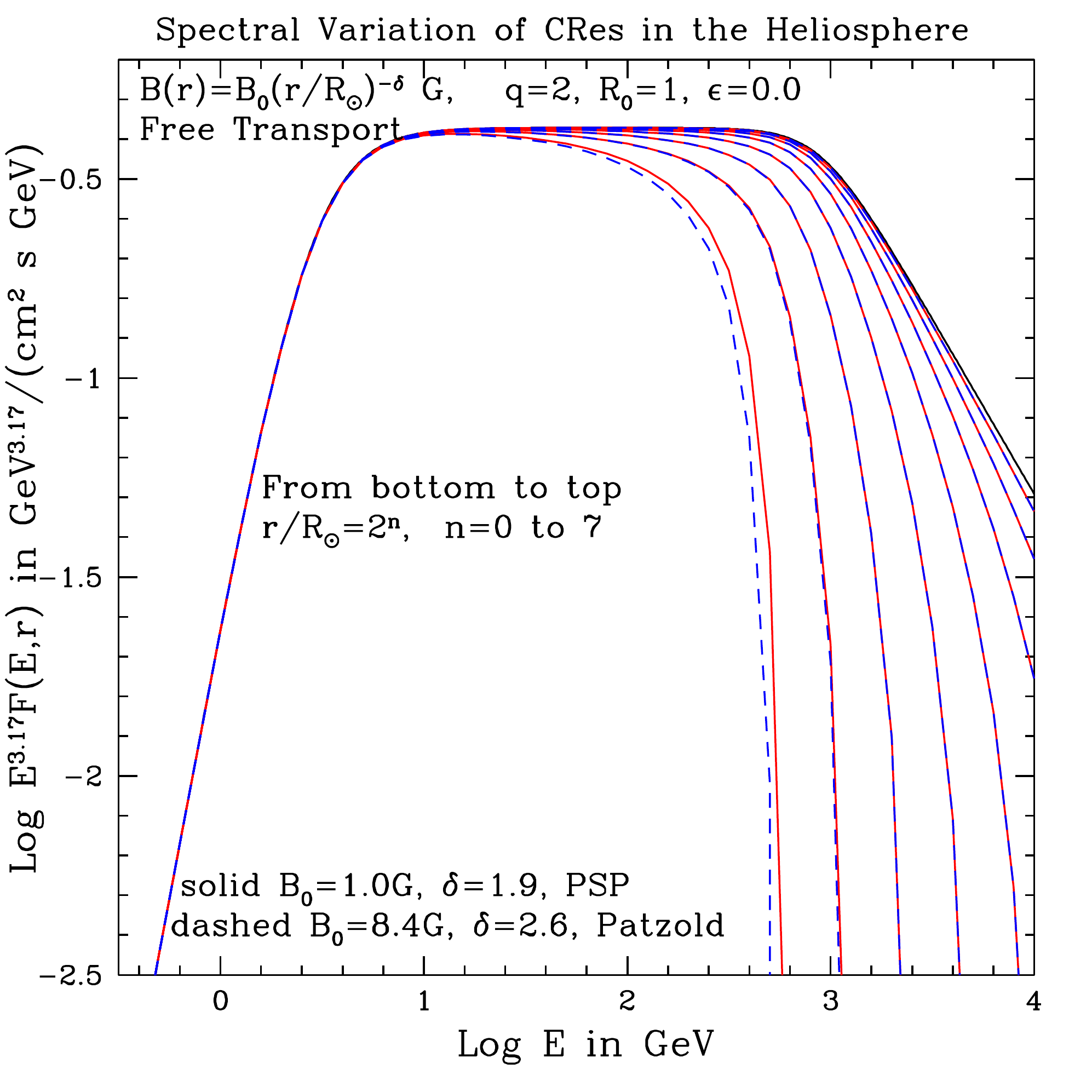}
			\includegraphics[width=2in]{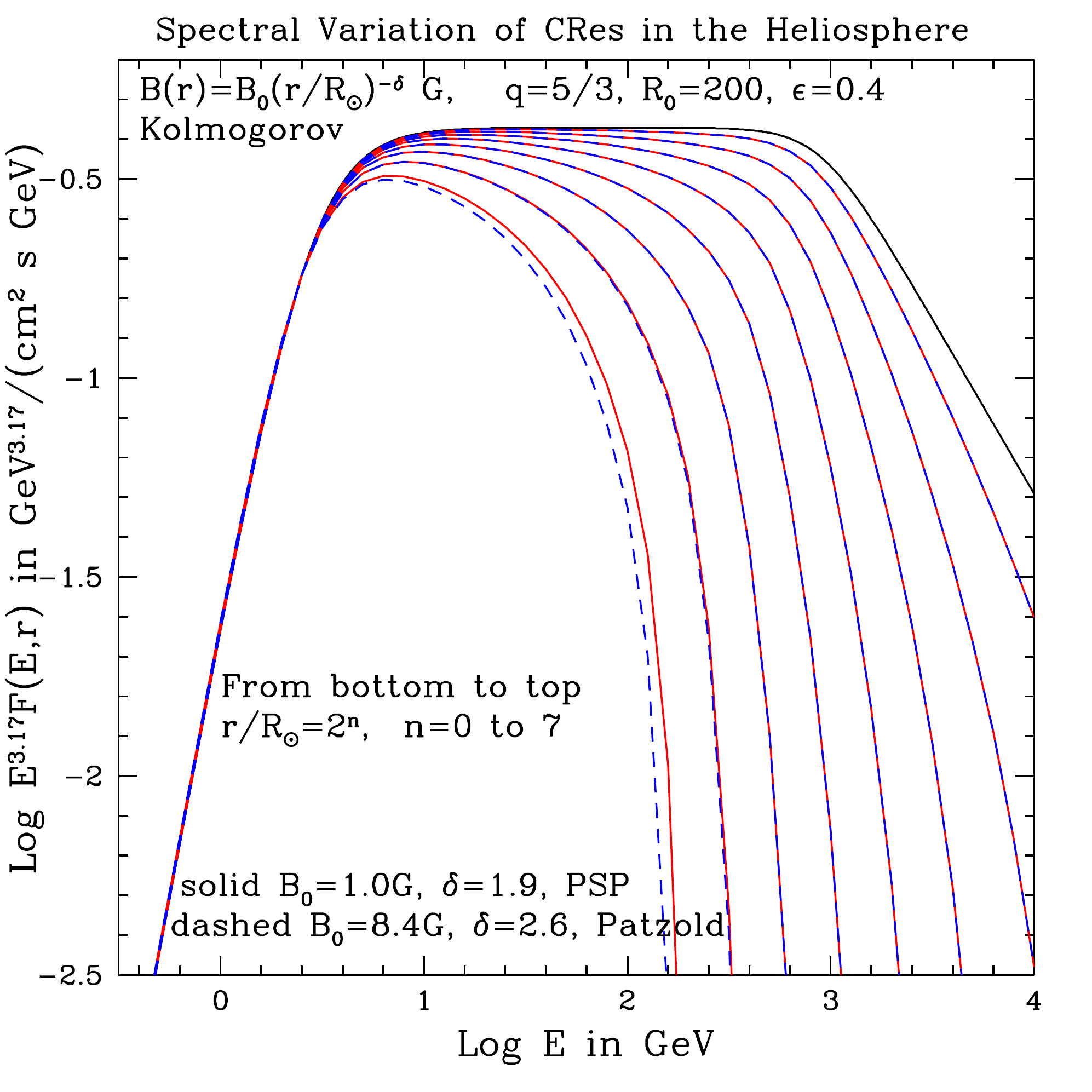}
			\includegraphics[width=2in]{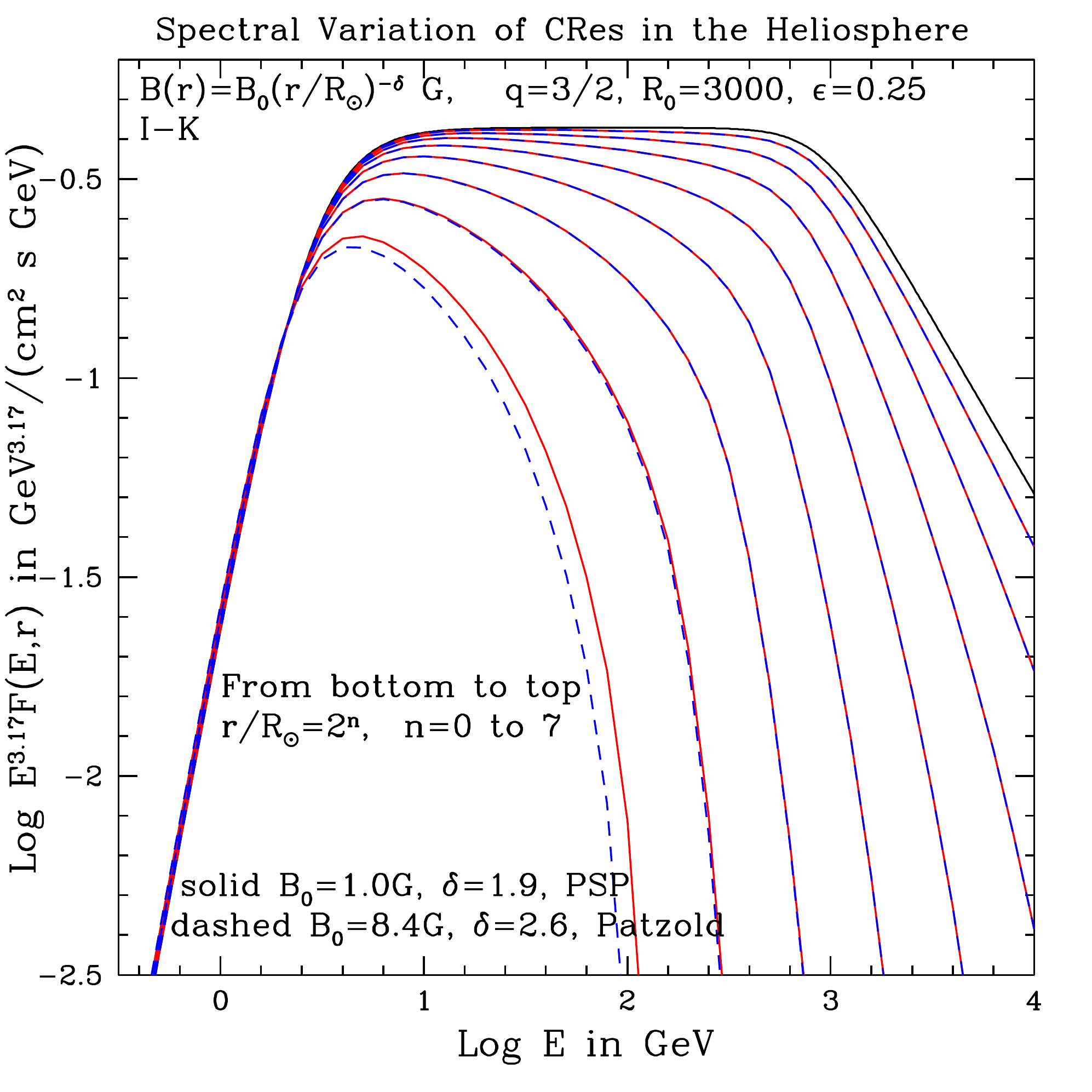}
			\caption{Variation with distance of the spectrum of Cosmic Ray electrons from 1 AU (black line fitted to observed  AMS02  and HESS points) to the Sun, at
				several  distances  obtained from  Equation (\ref{Nofr}). The top curve shows an analytic fit to the observed data described by Equation (\ref{CReSpec}). The red-solid  and blue-dashed
				curves show spectra for the two specified models of the $B$ field:
				Left: $q=2, R_0=1.0$ and $\epsilon=0.0$, which means we are ignoring scattering and field
				convergence effects.  Middle for Kolmogorov model with $q=5/3, R_0=200$ and $\epsilon=0.40$   and right for I-K model with $q=3/2, R_0=3000$ and  $\epsilon=0.25$; see Figure \ref{fig:Rofx}, left.}
			\label{fig:spectra}
		\end{center}
	\end{figure}

	In Figure \ref{fig:spectra} we show spatial variations of the CRe spectra flux, $F(\g, r)=cN(\g, r)$, from 1 AU to the
	photosphere%
	\footnote{As described in \S (3.1), the kinetic equation used here is only approximately true at high energies in the outer $r>r_{\rm cr}\sim 10\rs$ region. However, as seen in these figures most of the variation of  spectra occurs in the inner region where the kinetic approach is required.}
	for three models of the transport and two scenarios of the $B$ field structure (Patzold
	and PSP) described  in Equation
	(\ref{Bfits}). First we consider the {\it free transport case} ignoring scattering and field
	convergence, which means we set $q=2, R_0=1.0$ and $\epsilon=0$. The results are shown on the left
	panel of Figure \ref{fig:spectra}, which represent the minimum effect of the transport on spectral
	variation. We also show spectra for two more realistic models of turbulence; The I-K model with
	$q=3/2, R_0=3000, \epsilon=0.25$ in the middle panel, and Kolmogorov with $q=5/3, R_0=200,
	\epsilon=0.40$ in the right panel. 
	
\begin{figure}[!ht]
		\begin{center}
			\includegraphics[width=3in]{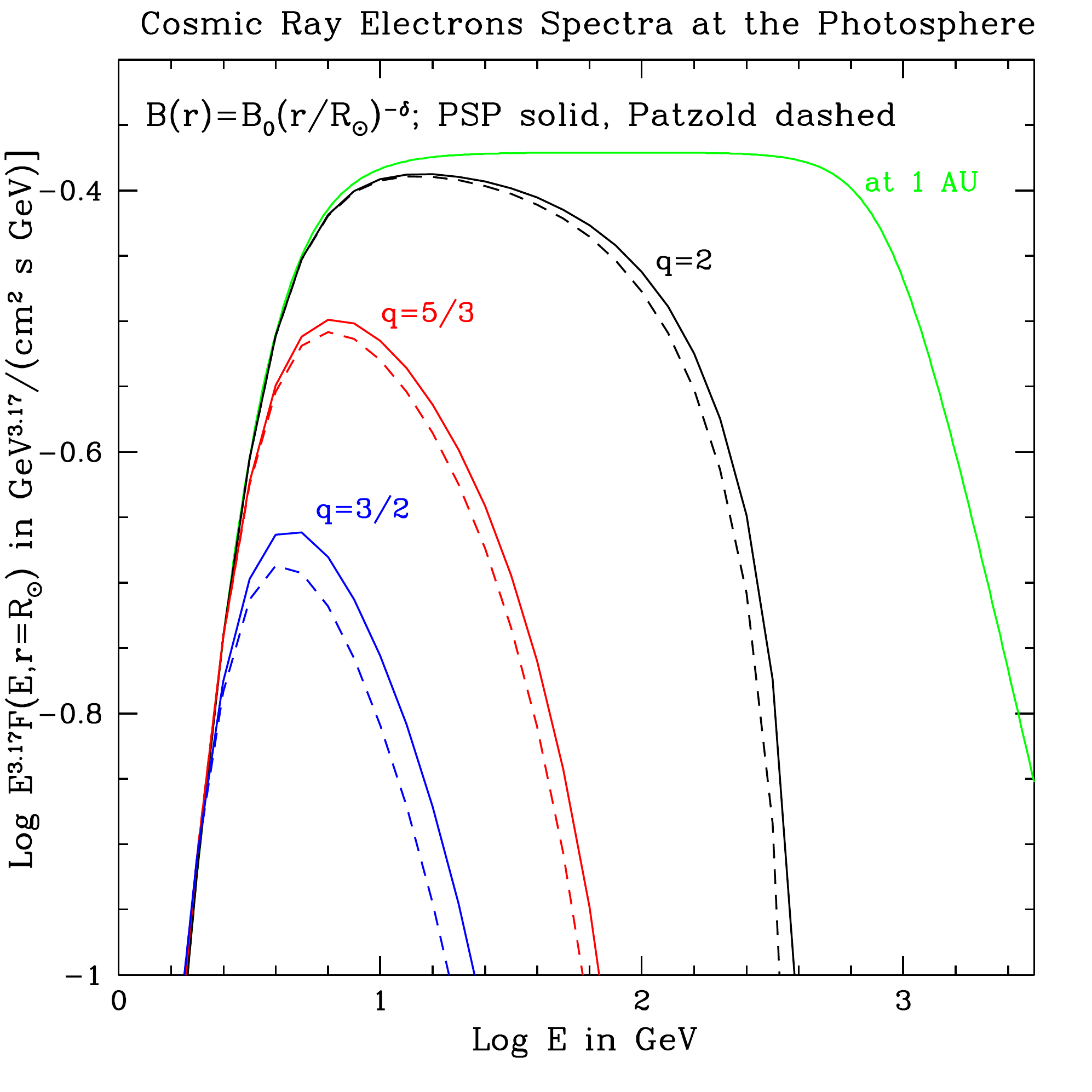}
			\includegraphics[width=3in]{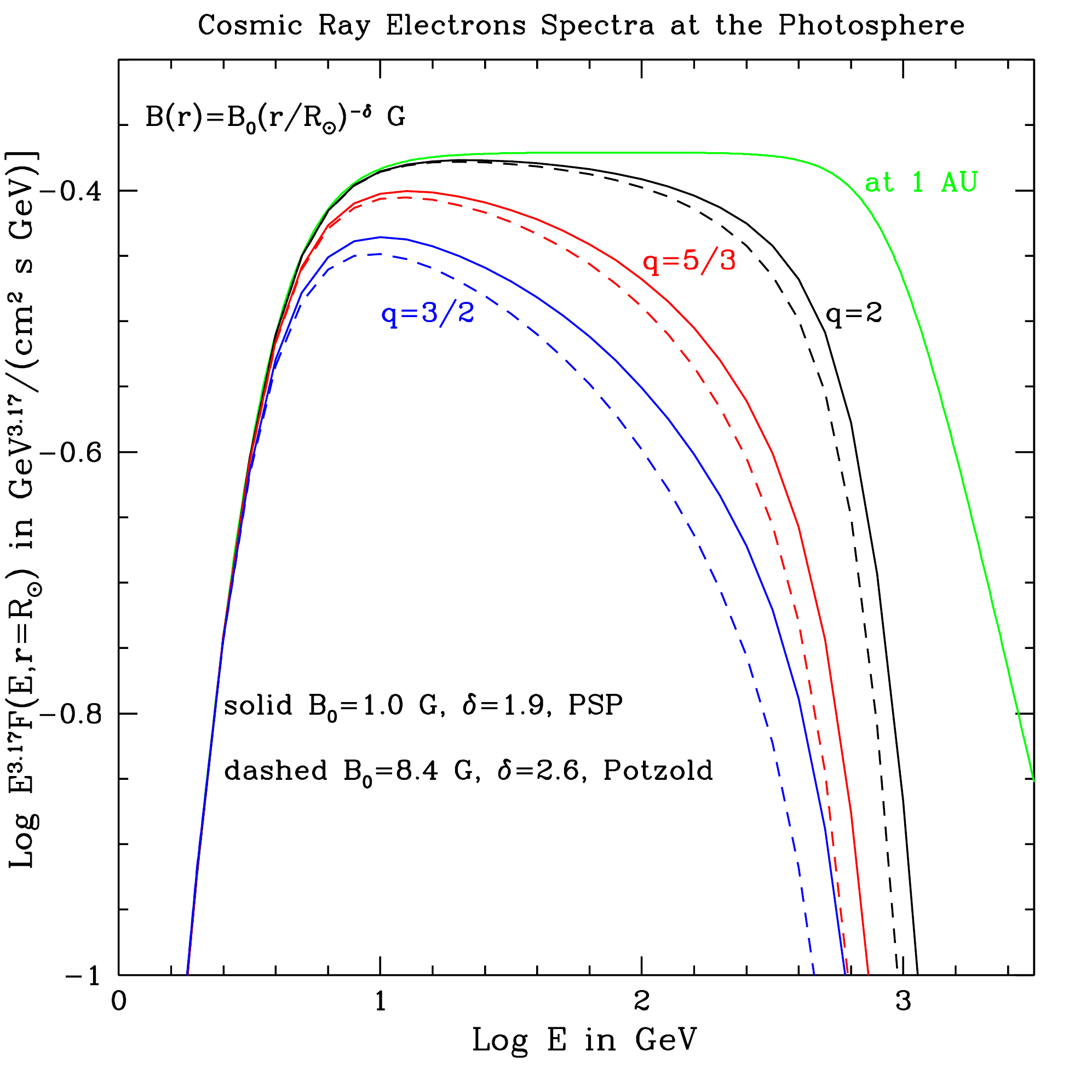}
			\caption{CRe spectra at the photosphere for two $B$ field models (PSP and Patzold). Left, for the three
				models of turbulence with characteristics given in Figure \ref{fig:spectra} and right, for different values of the critical parameter $R_0=2000, 130$ for $q=3/2$ and 5/3, respectively}.
			\label{fig:sunspec}
		\end{center}
	\end{figure}

	As evident the most
	pronounce loss and modification of spectra  occurs near the Sun. For a closer
	comparison of spectra for different turbulence models, on left panel of Figure \ref{fig:sunspec} we show, for two  models of the $B$ field, the spectra at the photosphere for the three models of turbulence  used in Figure \ref{fig:spectra}. On the right panel, we show  spectra for  smaller values of  the critical parameter, $R_0=2000$ and 130 for q=3/2 and 5/3, respectively, to demonstrate the possible range of spectra.

	It should be noted that there are some uncertainties in the value of the three primary model parameters, $B_0, \epsilon$ and $q$, used here, in addition to the uncertainty in the normalization values discussed above (Eq. \ref{N0ofr}).
	
	Substituting these electron spectra in Equation (\ref{emissivity}) one can calculate the spectra of synchrotron and IC
	emissivities, $\eta(\nu,r)$, as a function of distance from the Sun, for appropriate interaction cross sections, which  depend directly on
	the variations with $r$ of the $B$ field and optical photon  energy density, respectively.

	\begin{figure}[!ht]
		\begin{center}
			\includegraphics[width=6in]{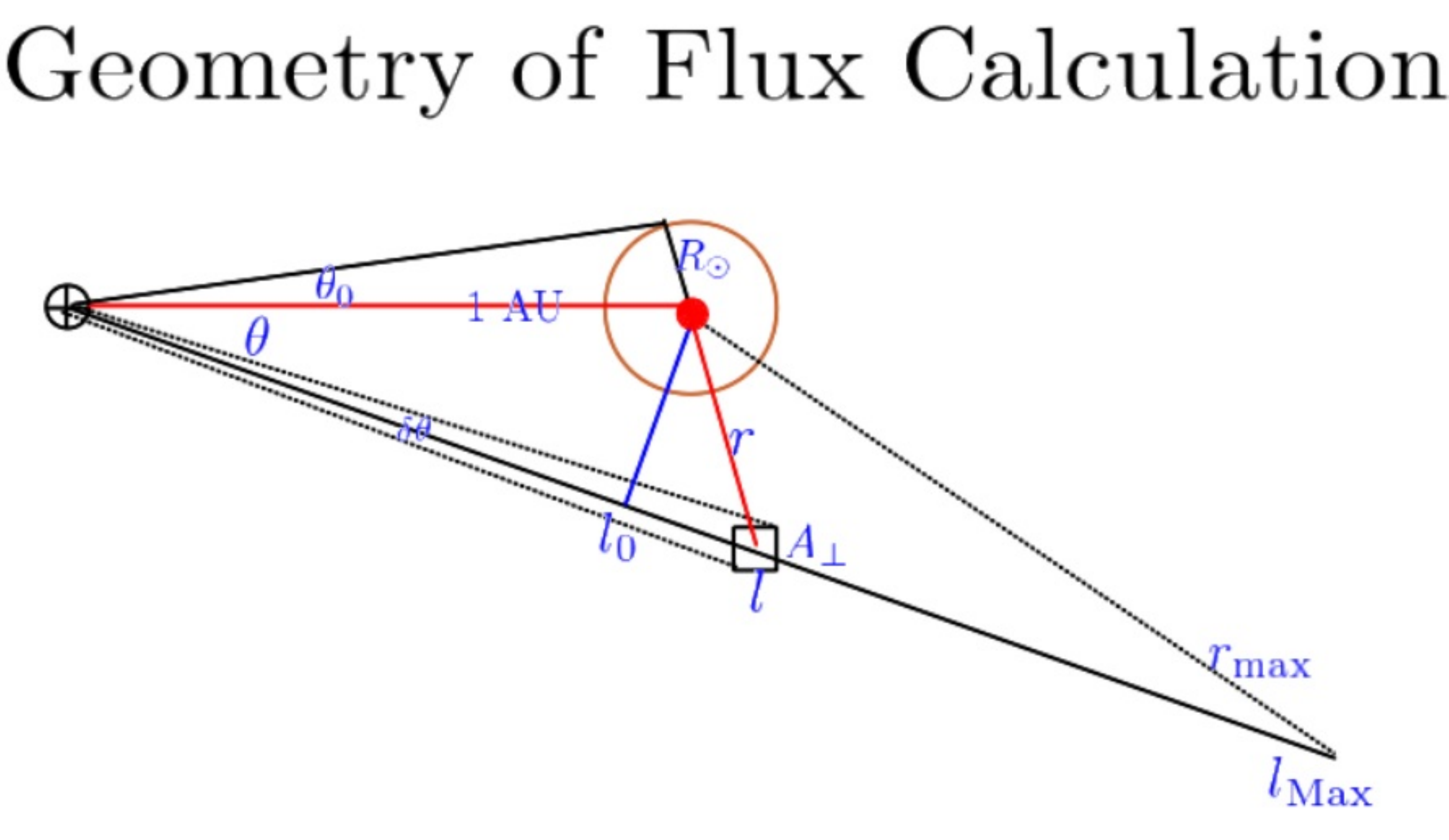}
			\caption{Geometry of calculation of the expected radiative flux at the Earth.}
			\label{fig:geom}
		\end{center}
	\end{figure}
	
	\subsection{Expected Radiative Flux  at 1 AU}
	\label{sec:FluxatAU}
	
	The observed flux of radiation at Earth at the distance 1 AU from the Sun will
	depend on
	the angle $\t$ between the observation line of sight and the Sun-Earth connection and angular area
	$d\Omega=\sin\t d\t d\phi$, depicted in Figure \ref{fig:geom}. The total flux will be an integral over
	the line of sight;
	
	\beq
	\label{flux1}
	F(\nu, \t)\d\Omega=\int_0^\infty \frac{\eta(\nu,r)(dV/dl)}{ 4\pi l^2}dl=\frac{d\Omega}{4\pi}\int_0^\infty
	\eta(\nu,r)dl,
	\eeq
	where the volume element $dV=A_\perp dl$ yielding $(dV/dl)/(4\pi l^2)=(d\Omega/4\pi)$, and 
	$r^2=a^2+l^2-2al\cos \t$, where $a=1$AU. From this
	we find $dl/dr=\pm r/\sqrt{r^2-a^2\sin^2\t}$, with the minus sign  for the integral from the Earth,
	$l=0$  to $l=l_0=a\cos\t$ (or $r=a$ to $r=a\sin\t=l_0\tan\t$), and the plus sign for $l=l_0$
	($r=a\sin\t$) to, in principal, infinity but in practice
	we can use $l_{\rm max}=2a\cos\t$  (or $r_{\rm max}=a$) as the upper limit, since most of the
	radiation  will come from 
	the vicinity of the Sun. This makes the two integrals equal  yielding 
	\beq
	\label{flux2}
	F(\nu, \t)\d\Omega=\frac{\d\Omega}{2\pi}\int_{a\sin\t}^a \eta(\nu,r)\frac{rdr}{\sqrt{r^2-a^2\sin\t^2}}.
	\eeq
	Thus, to calculate the expected radiation  fluxes from the solar disk and around it we need  the electron spectra,
	$N(\g,r)$ from 1 AU to the Sun.
	For the azimuthally symmetric situation at hand, we  integrate over $\phi$ to
	obtain $\d\Omega=2\pi \sin\t d\t$, and using the dimensionless  distances $x=r/\rs$,
	$x_{au}=a/\rs=214=1/\sin\t_0$, we then obtain
	
	\beq
	\label{flux3} 
	F(\nu, \t)\d\Omega=\rs(\sin\t d\t)\int_{x_{au}\sin\t}^{x_{au}}  \eta(\nu,x=r/\rs)\frac{xdx}{
		\sqrt{x^2-x^2_{au}\sin^2\t}}.
	\eeq
	
	\subsubsection{Flux From the Solar Disk and Beyond}
	\label{sec:diskF}
	
	For emission at the photosphere (i.e.~$\t\leq\t_0$), we see only half of the flux because of the high optical depth of the Sun, and the lower limit of the integral, $x_{au}\sin\t=1$ independent of $\t$.
	Thus, we can change the order of integration, first integrating over the angle $\t$ from 0 to $\t_0$ and obtain the  flux from the whole disk
	
	\beq
	\label{diskflux1}
	F_{\rm disc}(\nu)=(1/2)\rs\int_1^{x_{au}} \eta(\nu,x=r/\rs)xdx\int_0^{\t_0}\frac{\sin\t d\t}{
		\sqrt{x^2-x_{au}^2\sin^2\t}},
	\eeq
	Since $\t_0\ll 1$ we can set $\sin\t=\t$, in which case the integral over $\t$ can be carried out
	easily yielding 
	
	\beq 
	\label{diskflux2}
	F_{\rm disc}(\nu)=(1/2)\rs\sin^2\t_0\int_1^{a/\rs} \eta(\nu,x=r/\rs)x(x-\sqrt{x^2-1})dx,
	\eeq
	with $(1/2)\rs\sin^2\t_0=7.6\times 10^5$ cm. At the photosphere the emission increases by a factor of 2.%
	\footnote{For emission near the limb one must consider optical depth effects which depends on the energy of emitted photons. In general the optical depth decreases rapidly above the photosphere except at radio regime where synchrotron self absorption and free-free (or bremsstrahlung) absorption may remain significant to a larger distance.}
	For radiation from regions larger than the Sun, the angle
	integrated flux can be obtained as:
	\beq
	\label{lessthanthetaflux}
	F(\nu, <\t)=F_{\rm disc}(\nu)+\int_{\t_0}^\t F(\nu, \t')\sin'\t d\t',
	\eeq
	where $F(\nu, \t')$ is given in Equation (\ref{flux3}).
	
	In Summary, given the radiation emissivity $\eta(\nu, r)$ and spectral variation of CRe spectrum, Eq. (\ref{Nofr}), we
	can obtain the expected fluxes from
	the disk, Equation (\ref{diskflux2}), and areas larger than the disk from Equation (\ref{lessthanthetaflux}).
	
	\section{Summary and Conclusions}
	\label{sec:summary}
	
	CRs are observed in great details by the near Earth instruments around 1 AU and beyond in the outer heliosphere, but there are scant measurements of CRs in the inner heliosphere. However, CRs  interacting with solar gas and fields can produce high energy radiation mainly in the gamma-ray range (observed first  by EGRET instruments on board CGRO and in greater detail by the Large Area Telescope  on board {\it Fermi})  involving well known emission processes  during the quiet phases of the Sun (QS).  Interpretation of these emission processes requires a knowledge of  the flux, spectrum and other characteristics of CRs in the inner heliosphere (inside 1 AU), which requires an accurate treatment of the transport of CRs from 1 AU, where these characteristics are known, to the Sun.  
	
	The past interpretations of these radiation have treated the transport of CRs using a phenomenological modulation method (see, review by \citet{Orlando}), which has had some success treating transport of CRs from outer boundaries of the solar wind to 1 AU. The primary physical process affecting this transport is the interaction of CRs with turbulence in the solar wind, but in the inner heliosphere other effects such as presence of a strong guiding magnetic field and, for CRes, energy losses become dominant. The modulation methods do not treat these aspects. In particular,  to best of our knowledge, the important role of energy losses has not been treated quantitatively. 
	
	The aim of this paper is the development of an algorithm for this analysis with focus on the transport of CRes. It is well known that CRes interacting with solar optical photons near the Sun produce some of the observed gamma-rays via the IC process \citep{Abdo11}. Our results can provide a more accurate treatment of this radiation.  On the other hand, there has not been any estimate of possible synchrotron radiation by CRes spiraling along magnetic field lines. Our eventual goal is an accurate calculation of the synchrotron emission using the transport method we have developed in this work. Observation and interpretation of the synchrotron emission by CRes is best carried out during QS when, as we have shown, synchrotron emission near the Sun is expected to be mainly in the extreme ultraviolet (EUV) to hard X-ray (HXR) range. Observation by RHESSI \citep{Hannah} has provided a robust upper limit on the QS flux in the X-ray band from 3 to 100 keV, which can constrain the predictions of the synchrotron model.
	
	Below we give a brief summary of the salient aspects of our paper relevant to the transport of CRe from 1 AU to the Sun.
	
	\begin{enumerate}
		
		\item
		
		We show that treatment of transport requires {\it kinetic approach} because for prevailing $B$ fields, especially closer to the Sun,  the gyroradius of  > GeV electrons is smaller than the $B$ field scale height so that CRes are tied to the strong guiding field and spiral around them losing significant amount of energy (and producing synchrotron radiation). This transport can be treated by the Fokker-Planck equation including three main ingredients;  $B$ field convergence, synchrotron and IC energy losses, and scattering by turbulence.  The first two require structure of the amplitude magnetic field and variation of well known photon energy density. The last requires  spatial variation of the energy density and spectrum of turbulence.
		
		\item
		
		Several instruments, in particular PSP, provide in-situ measures of $B$ field  down to about 0.1 AU, and there are several indirect estimates of the $B$ field inside this region. We use a combination of these measurements described in \S \ref{sec:bfield}. For IC energy loss we use solar photon energy variation derived by \citet{Orlando2008}.  Several earlier in-situ measurements describe the characteristics of turbulence at around 1 AU and PSP has extended these measures to about 0.1 AU. We have no measurements below this so we rely on extrapolation and some theoretical calculations in this region.
		
		\item
		
		Instead of solving the kinetic equation numerically we use a simpler method based on numerical simulations by \citet{EP18}, which accounts for field convergence and pitch angel diffusion, $D_{\mu\mu}$, due to scattering with turbulence.  This method introduces  the concept of  resident time or escape time, which allows to separate  determination of the spatial and spectral variations.  In addition, the ratio of the escape to free crossing time enhances  the rates of synchrotron and IC energy losses, and reduces the overall density of the CRes. The density, however, increase toward the Sun, in  regions where gyro radius is small, due to focusing effect of the converging field lines.  This critical time scale, described in Equation (\ref{MKEP}), depends on the free advection or crossing time, $\tcross$, on $B$ field scale height, $H_B$, and the mean free path, $\lambda_{\rm mfp}$, or scattering time, $\tsc=\lambda_{\rm mfp}/v\sim \langle 1/D_{\mu\mu}\rangle$, which is the most critical scale.
		
		\item
		
		In general, the mean free path  or scattering time depend on characteristics of turbulence, plus $B$ field and plasma density, in a complicated way. However, for relativistic electrons, which interact mainly with low frequency Alfv\'en or fast mode turbulence, this relation is considerably simplified and depends on two parameters; the Alfv\'en speed and a interaction rate (or time scale) that bundles several turbulence and $B$ field parameters into one, described by Equation (\ref{taup}), which varies with distance in a complicated way. The detailed descriptions of variation with distance of turbulence characteristics and calculation of this critical time scale  are given in Appendices A and B. These results are summarized in Figures \ref{fig:Rofx} (left)  and \ref{fig:TurbParams}. The energy dependence in the relativistic regime is simple with $\tsc\propto \g^{2-q}$, where $q$ is the power law index of turbulence in the inertial range, and according to PSP measurement it varies from Kolmogorov (outer region) value of 5/3 to I-K value of 3/2, from about 0.3 to 0.2 AU. 
		
		\item
		
		We use both of the above  models and a third with $q=2$ to account for free (unaffected by $B$ field and turbulence) transport case, which shows the effects of energy loss alone,  and calculate the spectral and spatial variation with distance of the CRe spectrum and density (or flux) from their measured values at 1  AU to the Sun. The spatial variation derivation is detailed in Appendix D.
		\item
		
		These spectra can be used to calculate the emissivity of synchrotron and IC emissions, as described in \S \ref{sec:emissivities}, and the expected radiation flux at 1 AU from the disk of the Sun and regions around it, described in \S \ref{sec:FluxatAU}.
		
	\end{enumerate}
	
	In forthcoming papers, using these models of transport and resultant CRe spectra, we will calculate the IC spectra more accurately than done previously, and the synchrotron spectra for the first time. These can then be compared with observations by {\it Fermi}-LAT and RHESSI.

	
	The work of VP is supported by NASA Living With Star program grant NNH20ZDA001N-LWS. E.O. acknowledges the ASI-INAF agreement n. 2017-14-H.0, the NASA Grant No. 80NSSC20K1558.
	
	\section{Appendix A: In Situ Measurements of Turbulence}
	\label{sec:turbulrnce}
	
	In this section we summarize three in situ measurements of  intensity and spectrum of turbulence in
	the frequency range of $10^{-5}<f<1$ Hz, with corresponding  wave numbers $k=2\pi f /v_A$. Most of these  show a
	Kolmogorov spectrum, with power-law index $q=5/3$ in the inertial range ($f_{\rm min}<f<f_{\rm max}$). There is generally some steepening above, and most measurements
	show a distinct spectral flattening   below this range with index $q\sim 1$ down to measured  limit of  $f_{\rm lim}\sim 10^{-5}$ Hz. We need
	to calculate the total energy density of turbulence ${\cal W}_{\rm turb}=\int_{f_{\rm lim}}^{f_1}
	{\cal W}(f)df$.
	Assuming that to spectrum below $f_{\rm min}$ extends to $f_{\rm lim}$ with index $q=1$, it is easy to show
	that
	\beq
	\label{turbTotal}
	{\cal W}_{\rm turb}=f_{\rm min}{\cal W}(f_{\rm min})\frac{1}{ q-1}\left[1-(f_{\rm min}/f_{\rm max})^{(q-1)}+\ln (f_{\rm min}/f_{\rm lim})\right].
	\eeq
	
	Leamon et al. (1998. 1999) measure a Kolmogorov spectrum ($q=5/3)$ at $\sim 1$ AU with the specific energy density of
	${\cal W}(f_{\rm min})=10^3$ (nT)$^2$ Hz $^{-1}$, in the  inertial range $f_{\rm min}=10^{-3}, f_{\rm max}=0.1$ Hz. The spectrum
	steepens to  $q\sim 3$ above 0.1 Hz. This yields a total turbulence energy density of 
	${\cal W}(f>f_{\rm min})\sim 1.5 $(nT)$^2$. They do not provide any measurements below $f_{\rm min}$, but extending
	this to $f_{\rm lim}=10^{-5}$ yields ${\cal W}_{\rm turb}\sim 7$ (nT)$^2$. 
	
	Bruno and Carbone (2013) show the Helios measurements at $r=0.9, 0.7$ and 0.3 AU, with Kolmogorov
	spectrum between $f=0.1$ and $f_{\rm min}$, which decreases slightly with distance. Below $f_{\rm min}$ the spectrum
	flattens to $q\sim 1$ down to $f_{\rm lim}=2\times 10^{-5}$ Hz. From these we obtain
	\beq
	\label{Helios}
	{\cal W}(f>f_{\rm min})=
	\begin{cases}
		5.0 \,\, {\rm nT}^2 & f_{\rm min}=6\times 10^{-4} \, {\rm Hz}, \, r=0.9 {\rm AU}\\
		15.0  \,\, {\rm nT}^2 & f_{\rm min}=2\times 10^{-3} \, {\rm Hz}, \, r=0.7 {\rm AU}\\
		156  \,\, {\rm nT}^2 & f_{\rm min}=5\times 10^{-3} \, {\rm Hz}, \, r=0.3 {\rm AU}.
	\end{cases}
	\eeq 
	Extending to $f_{\rm lim}=10^{-5}$ Hz, we obtain the total turbulence energy densities
	of 
	\beq
	\label{Helios2}
	{\cal W}_{\rm turb}(r)=
	\begin{cases}
		20\,\, {\rm nT}^2 &  \, r=0.9 {\rm AU},\\
		61 \,\,  {\rm nT}^2 &  \, r=0.7 {\rm AU},\\ 
		620 \,\,  {\rm nT}^2 &  \, r=0.3 {\rm AU}.
	\end{cases}
	\eeq
	
\begin{figure}[!ht]
		\begin{center}
			\includegraphics[width=5in]{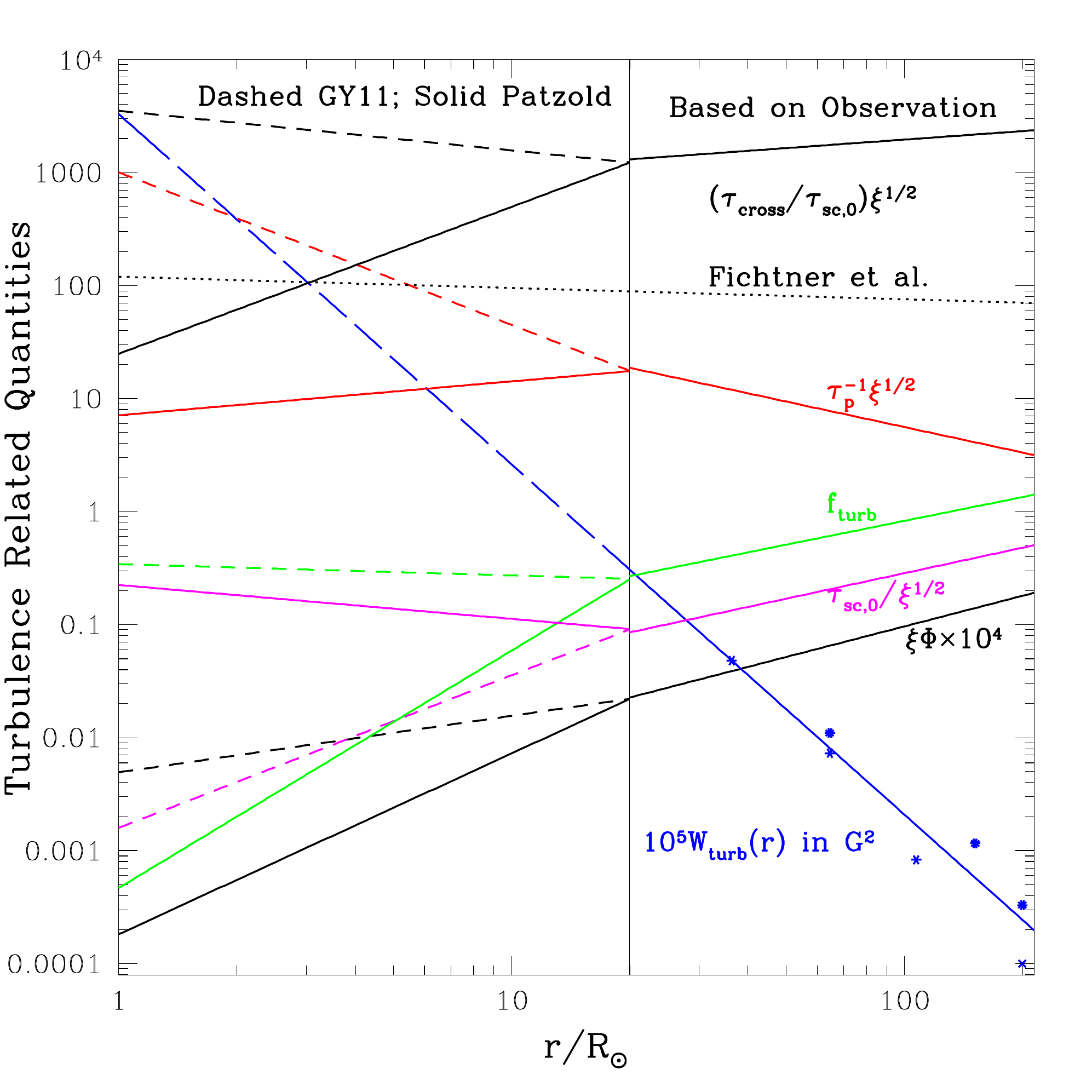}
			\caption{  
				Radial variations of CRe energy independent
				relevant quantities related to turbulence and scattering of the
				particles assuming the  I-K model with $q=3/2$.  $\Phi(x)$,
				Eq.
				(\ref{phi2}), (solid-black); turbulence energy density $W_{\rm turb}$, Eq. (\ref{FitW}), in blue;
				turbulence to $B$ field energy ratio, $f_{\rm turb}$, Eq. (\ref{fturb}), in green; characteristic
				turbulence rate $\tau_p^{-1}$, Eq. (\ref{taup1}), in red; scattering time $\tsc_{,0}$, Eq.
				(\ref{tscat1}), in magenta; and the crossing to scattering time ratio, Eq. (\ref{ratio}), in black.
				The results in the outer region are based on PSP measurements, with extrapolation to the inner part
				(long dashed). The two dashed and solid lines in the inner part are based on the two models of the
				$B$ field models, GY11 and Patzold. The dotted line is based on theoretical
				results from Fichtner et al. (2012).}
			\label{fig:TurbParams}
		\end{center}
	\end{figure}

	Recently, PSP  measurements (Chen et al. 2020)  have extended this information from  1 to 0.17 AU, and in the spectral range $2\times 10^{-5}<f<1$ Hz, showing a gradual change of the spectral
	index $q$ above $f_{\rm min}=10^{-3}$, from Kolmogorov 5/3 value at $r>0.3$ AU  to Iroshnikov-Kraichnin (I-K) value
	of 1.5 for $r<0.2$ AU. In a similar manner as above, estimation based on Figure 1 of Chen et al. (2020),  gives the following values for turbulence energy density.
	
	\beq
	\label{PSP}
	{\cal W}_{\rm turb}(r)=
	\begin{cases}
		10\,\,  {\rm nT}^2 &  f_{\rm min}=10^{-4} \, {\rm Hz},\,\,\,\,\,\,\,\,\,\,\, r=0.82 {\rm AU},\\
		83\,\,  {\rm nT}^2 &  f_{\rm min}=2\times 10^{-4} \, {\rm Hz},\,\, r=0.5 {\rm AU},\\
		730\,\,  {\rm nT}^2 &  f_{\rm min}=5\times 10^{-4} \, {\rm Hz},\,\, r=0.3 {\rm AU},\\
		4800\,\, {\rm nT}^2 &  f_{\rm min}=10^{-3} \, {\rm Hz},\,\,\,\,\,\,\,\,\,\,\, r=0.17 {\rm AU}.
	\end{cases}
	\eeq%
	The break frequency, $f_{\rm min}$, seem to increase inversely with distance as, $f_{\rm min}\sim 10^{-4}(AU/r)$\, Hz, with PSP showing slightly smaller values than Helios.
	The values of ${\cal W}_{\rm turb}(r)$ are plotted in Figure \ref{fig:TurbParams} showing  rough agreement
	between different estimates. A power-law fit to these 8 measured values yields 
	\beq
	\label{FitW}
	{\cal W}_{\rm turb}(r)= 12(r/{\rm AU})^{-3.1}\, {\rm nT}^2=0.033(r/\rs)^{-3.1} {\rm G}^2, \,\,\, 0.1<(r/{\rm AU})<1. 
	\eeq
	We will use this relation below.
	
	We have no information on turbulence energy density in the more critical inner region ($r<20\rs$). We consider two methods of extrapolation to the inner region. In one we assume that the above radial dependence continues to the Sun, then using the $B(r)$ field models described in \S 2 we calculate  the  ratio of turbulence to magnetic field energy densities, $f_{\rm turb}(r)={\cal W}_{\rm turb}(r)/[B(r)]^2$, needed for evaluation of $\tsc$ and ELEF (see below),  to be 
	\beq
	\label{fturb}
	f_{\rm turb}(r)=
	\begin{cases}
		0.033 (r/\rs)^{0.6} & \, 20 < r/\rs  < 214\\
		0.35 (r/\rs)^{-0.2} & \, r/\rs  < 20,\,\,\, {\rm GY11}\\
		4.7\times 10^{-4}(r/\rs)^{2.0} & \, r/\rs  < 20,\,\,\, {\rm Patzold}.
	\end{cases}
	\eeq
	However, considering that the turbulence energy density is almost proportional to $[B(r)]^2$ in the outer region, it is reasonable to assume that, like the $B$ field in Patzold model,  ${\cal W}_{\rm turb}(r)$ increases faster in the inner region yielding a flatter, nearly constant $f_{\rm turb}(r)\sim 0.3$. We will use a combination of both these extrapolations shown in Figure \ref{fig:Rofx}.

	\section{Appendix B: Scattering Time}
	\label{sec:scat2}
	
	The pitch angle diffusion coefficient, $D_{\mu\mu}$,  and hence $\kappa_{ss}$ and $\tsc$, can be
	obtained from  gyro-resonance interaction rates of particles with plasma waves of frequency $\omega$
	and wave vector $k$, obeying the resonance condition 
	\beq
	\label{res}
	\omega(k)-k_\|\mu v=\pm \Omega/\g,
	\eeq
	where $v$,  $\g$ and $\Omega$ are the velocity, Lorentz factor and gyrofrequency of the the
	particle, and $k_\|$ is the component of the wave vector parallel to the $B$ field.  The interaction
	rates depend on the dispersion relation of the waves, $\omega(k)$, the energy
	density of the waves, ${\cal W}_{\rm turb}$, its spectrum (mainly the spectral index $q$ in the
	inertial range; $k_{\rm min}<k<k_{\rm max}$), and the  background plasma $B$ field and
	density, $n$ (or Alfv\'en velocity, $v_A$).  However, for a power-law spectrum of turbulence, ${\cal W}(k)={\cal W}(k_{\rm min})(k/k_{\rm min})^{-q}$, the diffusion rate (or scattering time)  scales with the
	characteristic time $\tau_p$ or characteristic rate (see, e.g.~Dung \& Petrosian 1994) 
	\beq
	\label{taup1}
	\tau_p^{-1}=(\pi/2)\Omega f_{\rm turb}(q-1)\Phi^{(q-1)} \,\,\,\, {\rm with}\,\,\,\, \Phi=ck_{\rm
		min}/\Omega,
	\eeq
	where fraction of turbulence energy density, $f_{\rm turb}\sim (\delta B/B)^2$ is given in Appendix A.
	
	In general, the scattering time, in addition to this scaling, depends in a complicated
	way on  $q$ and Alfv\'en velocity (or  $\beta_A=v_A/c$), and on particle energy and pitch angle
	(see, e.g.~Pryadko \&
	Petrosian 1997, 1999, Petrosian \& Liu 2004 Jiang et al. 2009). However, for high  energy protons and relativistic
	electrons with Lorentz factors $\g> m_p/m_e$, i.e.~energies  greater than 1 GeV, which is the case
	for
	our problem, the main interactions are with Alf\'ven waves, with the dispersion relation
	$\omega=k_\| v_A$ for waves propagating parallel to the $B$ field and for $\omega<\Omega_p$, the proton gyrofrequency, and with fast mode waves with
	$\omega=kv_A$ and $k\ll 1$ for both parallel and perpendicular wave propagation. In
	this case the equations describing the above characteristics are simplified considerably, especially
	when $\b_A\ll 1$, which  is  the case here. As shown in Figure~1,  $v_A$ has nearly a constant value of $\sim 500$ km/s in the inner region ($r<20\rs$) and 
	decreases  with distance  to 30 km/s at 1 AU  as $v_A=30(AU/r)^{1.2}$. 
	
	As shown in Pryadko \& Petrosian (1997) and Petrosian \& Liu (2004) for parallel propagating waves, the pitch angle
	averaged scattering time appropriate for relativistic electrons with isotropic pitch angle
	distributions, and for $\beta_A \ll 1$,  is
	\beq
	\label{tscat1}
	\tsc=\g^{(2-q)}\tsc_{,0} \,\,\, {\rm with} \,\,\, \tsc_{,0}/\tau_p=
	\begin{cases}
		2[(2-q)(4-q)]^{-1} & \, q<2\\
		3/4-(1/2)\ln \beta_A & \, q=2.
	\end{cases}
	\eeq
	Using  $v_A=500$ km/s for $r<20\rs$, where most of the losses take place, we obtain
	
	\beq
	\label{tscat2}
	\tsc_/\tau_p=
	\begin{cases}
		1.6\g^{(1/2)} & \, q=3/2\\
		2.6\g^{(1/3)} & \, q=5/3\\
		3.9 & \, q=2.
	\end{cases}
	\eeq
	
	\subsection{Scaling Details}
	\label{scaledetails}
	
	To complete the calculation of the scattering time we need to specify the numerical value of
	$\tau_p$ and its variation with distance, which depends on $\Omega_e$ (or $B$ field), $f_{\rm turb}$,  
	and the somewhat unknown  $k_{\rm min}=2\pi f_{\rm min}/v_A$, the inverse of the largest scale of the turbulence. This length scale  is related to
		the correlation length of the injected turbulence and is expected to be a fraction, $\xi$, of the size of the region, here $\sim r$, defined as 
		$\xi=2\pi/(k_{\rm min}r)$. The correlation length at the base of the corona, $r=\rs$, is estimated to be $2\pi/k_{\rm min}\sim 10^9$ cm implying $\xi\sim 0.03$. The correlation length most probably increases with distance. PSP observations of decreasing $f_{\rm min}\sim 10^{-4}(AU/r)$ in the outer regions seem to agree with this. Using $f_{\rm min}\sim 2\times 10^{-3}$ and $v_A=500$\, km/s at $r=20\rs$ yields  $\xi\sim 0.04$. For now we will keep  $\xi$ as a free parameter. 
	
	Using the general  magnetic field model of $B(r)=B_0x^{-\d}$, with $x=r/\rs$ we obtain 
	
	\beq
	\label{phi2}
	\Phi\frac{ck_{\rm min}}{\Omega_e}= \Phi_0\frac{x^{\d-1}}{ B_0}=\Phi_0
	\begin{cases}
		1.0x^{0.9} & \,  20<x<214\\
		3.3x^{0.5}& \, 1<x<20 \,\,\, {\rm BY11}\\
		0.12x^{1.6} &\, 1<x<20 \,\,\, {\rm Patzold},
	\end{cases}
	\eeq
	with $\Phi_0=1.5\times 10^{-7}/\xi\sim 3.7\times 10^{-6}$.
	PSP observations indicate that the spectral index changes from 5/3 to 3/2 between 0.3 and 0.2 AU.
	We are not aware of any direct measurement of the index closer to the Sun where the energy loss rate
	is most significant. Thus, we will consider three values of $q= 3/2, 5/3$ and 2. 
	Now substituting the above values for  $\Phi, q, f_{\rm turb}$ and the magnetic field, in Equations
	(\ref{taup1}) and (\ref{tscat1}) we can calculate $\tau_p^{-1}$ and $\tsc$. For example, for $q=3/2$
	we obtain
	
	\beq
	\label{taup2}
	\tau_p^{-1}=(0.04/\xi)^{1/2}
	\begin{cases}
		1.4\times 10^2 (r/\rs)^{-0.70} & \, 20 < x  < 214\\
		4.8\times 10^3  (r/\rs)^{-1.25} & \, x  < 20,\,\,\, {\rm GY11}\\
		76(r/\rs)^{0.2} & \, x  < 20,\,\,\, {\rm Patzold},
	\end{cases}
	\eeq
	
	and from Equation (\ref{tscat1})  we obtain
	
	\beq
	\label{tscat3}
	\tsc_{,0}=(\xi/0.04)^{1/2}
	\begin{cases}
		1.8\times 10^{-2} (r/\rs)^{0.75} & \, 20 < r/\rs  < 214\\
		3.4\times 10^{-4}  (r/\rs)^{1.35} & \, r/\rs  < 20,\,\,\, {\rm GY11}\\
		2.1\times 10^{-2}(r/\rs)^{-0.3} & \, r/\rs  < 20,\,\,\, {\rm Patzold},
	\end{cases}
	\eeq
	
	which for $\tcross=5.6x$ (Eq. \ref{tcross}) gives  the critical ratio (for $q=3/2$)
	
	\beq
	\label{ratio}
	\tcross/\tsc_{,0}=(0.04/\xi)^{1/2}
	\begin{cases}
		1.6\times 10^2 (r/\rs)^{0.25} & \, 20 < r/\rs  < 214\\
		8.2\times 10^3 (r/\rs)^{-0.35} & \, r/\rs  < 20,\,\,\, {\rm GY11}\\
		1.3\times 10^2 (r/\rs)^{1.3} & \, r/\rs  < 20,\,\,\, {\rm Patzold}. 
	\end{cases}
	\eeq
	Similar expressions can be obtained for the other two values of $q$. Left panel of Figure \ref{fig:Rofx} shows $\tcross/\tsc_{,0}$ for the three values of $q$. As evident there is a large difference between the two models of the $B$ field in the inner region. This is due to the extrapolation of ${\cal W}_{\rm turb}$ from outer to inner region, which as mentioned in Appendix A may not be correct. Using the flatter variation of $f_{\rm turb}$ in the inner region we obtain the dashed lines which are half way between the two models of $B(r)$. We will use these extrapolations, which give the ratio $\tcross/\tsc$ at $\g=1$. 
	
	The final crucial ratio $R(r, \g)$  can then be obtained using Equation (\ref{MKEP}).
	However, as evident $\tsc< \tcross$, even at the highest energies, $\g<10^6$, especially for $q=3/2$ and 5/3, implying that we are in the strong diffusion limit with  $R(r, \g)=[\tcross(r)/\tsc_{,0}(r)]\g^{(q-2)}$.
	
	There have been many theoretical attempts to estimate the radial and energy dependence of the above
	characteristics of turbulence in the heliosphere, in particular that of the scattering time or mean
	free path $\lambda_{\rm mfp}(r,\g)=\lambda_0(r/AU)^{\d_l}\g^{(2-q)}$. For example,  Chhiber et al.
	(2017) present several results from MHD simulations on the spatial variation (for $r<1$ AU) of the
	mean free path of protons with $0<\d_l<0.6$,  and $\lambda_0\sim 0.2 \pm 0.1$ AU
	or $\tsc_{,0}\sim 100$ s at 1 AU, which is much larger than $\tsc_{,0}$ 
	obtained from observations shown above. Fichtner et al. (2012) give result
	for
	scattering of 10 MeV protons by Alf\'ven waves showing $\lambda_{\rm mfp}/r$ varying from 0.02 at
	the Sun to 0.01 at 1 AU.  Based on results from Petrosian \& Liu (2004) this indicates a
	$\lambda_{\rm mfp}/r=0.02\g^{(2-q)}$ for relativistic electrons, or  $\tsc_{,0}\sim 0.047$ s at
	$r=\rs$.  
	
	The ratio $\tcross/\tsc_{,0}$ for  the Fichtner model is also
	shown on the left panel of  Figure \ref{fig:Rofx},  which lies between $q=3/2$ and 5/3 cases.  The actual effect of the transport coefficients is better
	demonstrated by the term $A(x, \g)$ defined in Equation (\ref{Aofx}) shown in Figure \ref{fig:Rofx}
	(right) for the three values of $q$ and Lorentz factors $\g=10^6, 10^5$ and $10^4$.
	
	As evident the spatial variation of this crucial factor is similar for the three values of the index $q$, but there is significant  differences in their energy dependencies. However, these differences are
	much smaller For $10^3<\g<10^6$ compared to the values shown for $\g=1$ (left panel). 
	
	\section{APPENDIX C: Perpendicular Diffusion}
	\label{sec:AppendixC}
	
	In strong guiding $B$ fields diffusion perpendicular to the field lines can be ignored, but when scattering mean free path becomes comparable to or larger than the particle gyro-radius,
	perpendicular diffusion becomes important. As shown in \S 3.1 the gyro-radius of relativistic electrons is $r_g=1.3\times 10^3\g({\rm G}/B)$ cm and the mean free path $\lambda_{\rm mfp}=c\tsc=ca(q)\g^{(2-q)}\tau_p$, where $a(q)$ and $\tau_p$ are given in Appendix B. Simple algebra  shows that
	\beq
	\label{gyro2}
	R_{\rm diff}\equiv r_g/\lambda_{\rm mfp}=\pi/[4a(q)]f_{\rm turb}(\g\Phi)^{(q-1)}.
	\eeq
	
	In the outer region ($r>20\rs$) with $q=5/3, a(q)=2.6, \Phi=\Phi_0x^{0.9}$ and $f_{\rm turb}=0.033x^{0.6}$ we obtain 
	\beq
	\label{gyroKol}
	R_{\rm diff}=1.3\times 10^{-7}\g^{2/3}x^{1.2}
	\eeq
	so that for $\g\leq 10^6$ $R_{\rm diff}\leq 0.8$ and 0.05 for at 1 AU and $r=20\rs$, respectively.
	In the inner region ($r<20\rs$) with $q=3/2, a(q)=1.6, \Phi=1.2 \Phi_0x^{1.6}$ and $f_{\rm turb}=4.7\times x^{2.0}$ we obtain
	\beq
	\label{gyroI-K}
	R_{\rm diff}=5\times 10^{-7}\g^{1/2}x^{2.6}
	\eeq
	so that for $\g\leq 10^6$, $R_{\rm diff}\leq 10^{-4}$ and 0.05  at the Sun and $r=20\rs$, respectively.
	
	Because we are mainly interested in the inner region the neglect of perpendicular diffusion is well justified.
	
	\section{APPENDIX D: Details of Density Variation}
	\label{AppendixD}
	
	Here we use the steady state, $\partial N/\partial t=0$, Equation (\ref{FPiso}). We first note that for relativistic CRes,  $\kappa_{ss}=3v^2\tsc=3c^2\tsc_{,0}\g^{(2-q)}$, so that if we multiply this equation by $q^{(q-2)}$, the first term in this equation will depend only on the spatial variable and describes the implicit spatial variation  $N_0(r)$. This also will alter the energy loss rate by this factor similar to what was used in \S \ref{sec:specvar}. If we now ignore the energy terms,  we can obtain a simple approximate solution for the CRe density variation by integration of the resultant equation over the volume, $dV(s)=A_\perp(s)ds$, of a bundle of field lines with cross section area, $A_\perp(s)$,  from the starting point $s=0$, the point where  kinetic equation becomes valid ($r=r_{\rm cr}\sim 10\rs$), to any $s$ (or any $r<r_{\rm cr}$). We set $ds=1.2dr, \tcross=1.2r/c$  and $R_0(r)=\tcross/\tsc_{,0}=R_0(r/\rs)^\epsilon$ to obtain $N(r)$ by integration over $s$ as:
	\beq 
	\label{Nesc1}
	\g^{(q-2)}\int_0^s A_\perp(s)\frac{\partial}{\partial s}\kappa_{ss}\frac{\partial N}{\partial s}ds=\frac{3c^2}{1.2}\left(\tsc_{,0} A\frac{dN}{dr}\right)_{r_{\rm c}}^r-\frac{3c^2}{1.2} \int_{r_{\rm cr}}^r N(r)\frac{d(\tsc_{,0}
		A)}{dr}=0.
	\eeq
	
	Since $A(r)\propto r^2$ and $\tsc_0(r)\propto r^{\d_{sc}}$ obey simple power laws we expect a power law behavior for $N(r)$ and can set $dN/dr=(N/r)(d\ln N/d \ln r)$, with the power law index $\d_N = d\ln N/d \ln r$ nearly a constant. This then allows to complete the integration, 
	which after some algebra leads to
	
	\beq
	\label{Nesc2}
	N(r)=N(r_{\rm cr})\left(\frac{A(r_{\rm cr})R_0(r)}{ A(r)R_0(r_{\rm cr}))}\right),
	\eeq
	which leads to the conjecture in Equation (\ref{Nofr}).


\begin{thebibliography}{}
		
		\bibitem[Abdo et al. (2011)]{Abdo11} Abdo, A. A., Ackerman, M. Ajello, M.,  et al.\ 2011, \apj, 734, 116
		\bibitem[Aguilar et al. (2014)]{AMS02} Aguilar, M., Aisa, D., Alvino, A., et al.,\ 2014 Phys. Rev. Letters, 113, 121102
		\bibitem[Alissandrakis \& Gary (2021)]{AG21} Alissandrakis, C. E., \& Gary, D. E.,\ 2021, {\it Front. Asron. and Space Sci.}, 7, 77A
		\bibitem[Badman et al. (2021)]{Badman} Badman, S. T., Bale, S. D., Rouillard, A. P., et al.,\ 2021, Astronomy \& Astrophysics, 650, 26
		\bibitem[Bruno \& Carbone (2005)]{Bruno} Bruno, R. \& Carbone, V.,\ 2013, {\it Living Review of Solar Physics}, 10, 2
		\bibitem[Chen et al. (2020)]{Chen} Chen, C. H. K., Bale, S. D., Bonnel, J. W., et al.,\ 2020, \apj\  Suppliment Series, 240, 53
		\bibitem[Chhiber et al.(2017)]{Chhiber} Chhiber, R., Subedi, P., Usmanov, A. V., et al.,\ 2017, \apj\ Suppliment Series, 230, 21
		\bibitem[Dung \& Petrosian (1994)]{DP94} Dung, R., \& Petrosian, V.,\ 1994, \apj, 421, 550
		\bibitem[Effenberger \& Petrosian (2018)]{EP18} Effenberger, F., \& Petrosian, V.,\ 2018, \apjl, 868, L28
		\bibitem[Fujii \& McDonald (2005)]{Fujii} Fujii, Z., \& McDonald, F. B.,\ 2005, Adv. in Sp Research, 35, 611
		\bibitem[Fichtner, et al. (2002)]{Fichtner} Fichtner et al., \ 2002, Astrtoparticle Physics, 17, 199
		\bibitem[Gopalswamy \& Yashiro (2011)]{GY11} Gopalswamy, Nat \& Yashiro, Seiji,\ 2011, \apjl, 736, L17
		\bibitem[Hannah et al. (2010)]{Hannah} Hannah, I. G., Hudson H. S., Hurford, G. J., \& Lin, R. P.,\ 2010, \apj, 724, 487
		\bibitem[HESS Collaboration (2017)]{HESS} H.E.S.S. Collaboration; Kerszberg, D., et al.,\ 2017, Presented at the ICRC, Buson, Korea
		640, L155 (2006), arXiv:astro-ph/0511149 [astro-ph].
		\bibitem[Hooper et al. (2017)]{Hooper} Hooper, D., Cholis, I., Linden T., \& Fang, K.,\ 2017, Phys. Rev. D 96, 103013.
		\bibitem[Jiang et al. (2009)]{Jiang} Jiang Y., Liu, S., \&  Petrosian, V.,\ 2009, \apj, 698, 163
		\bibitem[Leamon et al. (1998)]{Leamon98} Leamon, R. J., Smith, C. W., \& Ness,  N. F.,\ 1998, \jgr, 103, 4775
		\bibitem[Leamon et al. (1999)]{Leamon99} Leamon, R. J., Smith, C. W., Ness,  N. F.,\ 1999, \jgr, 104, 22331
		\bibitem[Lehblanc et al. (1998)]{Lehblanc} Lehblanc, Y., Dulk, G. A., Bougeret, J.-L.,\ 1998,  Sol. Phys., 183, 165
		\bibitem[Malyshkin \$ Kulsrud (2001)]{MK01} Malyshkin, L., \& Kulsrud, R.,\ 2011, \apj, 549, 402
		\bibitem[Moderski et al.(2005)]{mod05} Moderski, R., Sikora, M., Coppi, P.~S.,
		\& Aharonian, F.\ 2005, \mnras, 363, 954
		\bibitem[Moskalenko, et al. (2006)]{Moskalenko} Moskalenko, I. V., Porter, T. A., \& Digel, S. W.,\ 2006, \apjl, 652, L65
		\bibitem[Orlando \& Strong (2007)]{Orlando2007} Orlando, E. \& Strong, A.~W.\ 2007, \apss, 309, 359. doi:10.1007/s10509-007-9457-0
		\bibitem[Orlando \& Strong (2008)]{Orlando2008} Orlando, E., \& Strong, A. W.,\ 2008, A\&A, 480, 847
		\bibitem[Orlando \& Strong (2021)]{Orlando} Orlando, E., \& Strong, A. W.,\ 2021,  Journal of Cosmology and Astroparticle Physics, 04, 004
		\bibitem[Orlando (2008)]{OrlandoThesis}  Orlando, E. 2008, PhD Thesis, https://ui.adsabs.harvard.edu/abs/2008PhDThesis
		
		\bibitem[Petrosian(2012)]{VP85} Petrosian, V.,\ 1985, \apj, 299, 987 (Eq. A4)
		\bibitem[Petrosian(2012)]{VP12} Petrosian, V.,\ 2012, Space Sci Rev., 173, 535
		\bibitem[Petrosian \& Liu.(2004)]{PL04}  Petrosian, V., \& Liu, S.,\ 2004, \apj, 610, 550
		\bibitem[Patzold et al. (1987)]{Patzold} Patzold, M., Bird, M. K., Levy, G. S., et al.,\ 1987,  Sol. Phys., 109, 91
		\bibitem[Pryadko \& Petrosian (1997)]{PP97}  Pryadko, J.~M., \& Petrosian, V.,\ 1997, \apj, 482, 774
		\bibitem[Pryadko \& Petrosian (1999)]{PP99}  Pryadko, J.~M., \& Petrosian, V.,\ 1999, \apj, 515, 873
		\bibitem[Rybicki \& Lightman (1980)]{RL} Rybicki, G. B.  \& Lightman, A. P.,\ 1980 {\it Radiation Processes in Astrophysics}, J. Willy \& Sons, New York
		\bibitem[Saito et al. (1997)]{Saito} Saito, K., Poland, A. I., Munro, R. H.,\ 1977, Sol. Phys., 55, 121
	\end{thebibliography}
\end{document}